\documentclass[aps,prl,groupedaddress,twocolumn,superscriptaddress,amsfonts,amssymb,amsmath,showpacs,floatfix,preprintnumbers]{revtex4-2}

\usepackage{dcolumn}
\usepackage{bm}
\usepackage{multirow}
\usepackage{verbatim}
\usepackage{graphicx}
\usepackage{ulem}
\usepackage{color}
\usepackage{placeins}
\usepackage{xcolor}
\usepackage{hyperref}
\usepackage{slashed}
\usepackage{cleveref}

\begin{document}


\title{Inverse problem in the LaMET framework}

\author{Herv{\'e} Dutrieux}
\email[e-mail: ]{herve.dutrieux@cpt.univ-mrs.fr}
\affiliation{Aix Marseille Univ, Universit\'e de Toulon, CNRS, CPT, Marseille, France.}
\author{Joe Karpie}
\email[e-mail: ]{jkarpie@jlab.org}
\affiliation{Thomas Jefferson National Accelerator Facility, Newport News, VA 23606, U.S.A.}
\author{Christopher J.~Monahan}
\email[e-mail: ]{cjmonahan@coloradocollege.edu}
\affiliation{Department of Physics, Colorado College, Colorado Springs, CO 80903, U.S.A.}
\author{Kostas Orginos}
\email[e-mail: ]{kostas@jlab.org}
\affiliation{Physics Department, William \& Mary, Williamsburg, VA 23187, U.S.A.} 
\affiliation{Thomas Jefferson National Accelerator Facility, Newport News, VA 23606, U.S.A.}
\author{Anatoly Radyushkin}
\email[e-mail: ]{radyush@jlab.org}
\affiliation{Old Dominion University, Norfolk, Virginia 23529 U.S.A.}\affiliation{Thomas Jefferson National Accelerator Facility, Newport News, VA 23606, U.S.A.}
\author{David Richards}
\email[e-mail: ]{dgr@jlab.org}
\affiliation{Thomas Jefferson National Accelerator Facility, Newport News, VA 23606, U.S.A.}
\author{Savvas Zafeiropoulos}
\email[e-mail: ]{savvas.zafeiropoulos@cpt.univ-mrs.fr }
\affiliation{Aix Marseille Univ, Universit\'e de Toulon, CNRS, CPT, Marseille, France.}

\preprint{JLAB-THY-25-4295}

\begin{abstract}
One proposal to compute parton distributions from first principles is the large momentum effective theory (LaMET), which requires the Fourier transform of matrix elements computed non-perturbatively. Lattice quantum chromodynamics (QCD) provides calculations of these matrix elements over a finite range of Fourier harmonics that are often noisy or unreliable in the largest computed harmonics. It has been suggested that enforcing an exponential decay of the missing harmonics helps alleviate this issue. Using non-perturbative data, we show that the uncertainty introduced by this inverse problem in a realistic setup remains significant without very restrictive assumptions, and that the importance of the exact asymptotic behavior is minimal for values of $x$ where the framework is currently applicable. We show that the crux of the inverse problem lies in harmonics of the order of $\lambda=zP_z \sim 5-15$, where the signal in the lattice data is often barely existent in current studies, and the asymptotic behavior is not firmly established. We stress the need for more sophisticated techniques to account for this inverse problem, whether in the LaMET or related frameworks like the short-distance factorization. We also address a misconception that, with available lattice methods, the LaMET framework allows a ``direct'' computation of the $x$-dependence, whereas the alternative short-distance factorization only gives access to moments or fits of the $x$-dependence.
\end{abstract}

\maketitle

\section{Introduction}

In large momentum effective theory (LaMET) \cite{Ji:2013dva, Ji:2014gla}, the quasi-parton distribution function (quasi-PDF) $f(y, P_z)$ is constructed as the Fourier transform of matrix elements of operators with equal-time spacelike separation, denoted symbolically by $\mathcal{O}(z)$:
\begin{equation}
f(y, P_z) = \int_{-\infty}^{+\infty} \frac{\mathrm{d}z P_z}{2\pi}\,e^{i y z P_z} \langle P | \mathcal{O}(z) | P \rangle\,, \label{eq:quasiPDF}
\end{equation}
where $z^\mu = (0, 0, 0, z)$. The quasi-PDF is related to the renormalized light-cone PDF, $q(x, \mu^2)$, by a matching procedure with coefficients computable in perturbation theory~\cite{Izubuchi:2018srq,Chen:2020ody,Li:2020xml}:
\begin{equation}
q(x, \mu) = \int_{-\infty}^{+\infty} \frac{\mathrm{d}y}{|y|}\,C\left(\frac{x}{y}; \frac{\mu}{x P_z}\right) f(y, P_z)\,, \label{eq:matching}
\end{equation}
up to power corrections that vary as $\mathcal{O}(\Lambda^2 / x^2 (1-x) P_z^2)$, where $\Lambda$ is a typical hadronic energy scale~\cite{Braun:2018brg,Zhang:2023bxs}. A large momentum $P_z$ is therefore 
necessary to reduce the size of power corrections, increase the reliability of the perturbative matching of Eq.~\eqref{eq:matching}, and increase the accessible range of Fourier harmonics $\lambda = z P_z$ in Eq.~\eqref{eq:quasiPDF}.

However, a drawback of large momenta in lattice QCD is the exponential loss of signal and the exponentially increasing difficulty of isolating the ground state from hadronic excitations. While improved operators can enhance the ground state signal~\cite{Bulava:2011yz,Aubin:2011zz,Bali:2016lva,Fischer:2020bgv,Wagman:2024rid,Hackett:2024xnx,Chakraborty:2024scw,Zhang:2025hyo}, the exponential growth in variance seems to forbid momenta significantly larger than 3 GeV for the foreseeable future with current methods. 

In addition, with current renormalization strategies, including the ratio method~\cite{Musch:2010ka,Orginos:2017kos}, RI-MOM~\cite{Alexandrou:2017huk,Zhang:2020rsx}, hybrid~\cite{Ji:2020brr}, and self-renormalization~\cite{LatticePartonLPC:2021gpi}, noise increases significantly when the length of the Wilson line $z$ increases. This can be traced to the increase of noise caused by the ratio of exponentially shrinking noisy numbers. 

As a result, the available range of Fourier harmonics $\lambda = z P_z$ is quite limited, and the largest Fourier harmonics are often noisy, if not simply unreliable. In fact, state-of-the-art calculations involving a momentum $P_z$ of the order of 2 GeV and reasonably light quark masses often lose signal by $\lambda = z P_z$ of the order of $5 - 8$ (for a far from exhaustive set of examples, see~\cite{Zhang:2020dkn,Alexandrou:2021oih,Gao:2021dbh,HadStruc:2021qdf,Fan:2022kcb,Gao:2023ktu,Delmar:2023agv, Ding:2024saz,Good:2024iur}). The construction of $f(y, P_z)$ from such a limited range of noisy and possibly unreliable data is therefore an inverse problem worthy of particular attention. For the standard definition of ``inverse problem'', see, for example, the preface of Ref.~\cite{doi:10.1137/1.9780898717921} and Ch.~1 of Ref.~\cite{Aster2013}.

At very large distances $z$, the renormalized correlation function decays exponentially. At asymptotic distances, a nonzero contribution is generated by a meson propagating along the spacelike distance $z$, described by a Bessel function that decays at large $|z|$ exponentially with the pion mass~\cite{Gao:2021dbh}. At intermediate distances, multi-particle or excited states could also contribute, and a different effective mass for the exponential may be appropriate, especially when considering physical quark masses. The expected exponential decay with $z$ is therefore given by a distance $r = m_{\textrm{eff}}^{-1}$ of the order of 1 fm. For $P_z = 2$ GeV, this corresponds approximately to an exponential decay in the Fourier harmonics of $\lambda_r \equiv r P_z \simeq 10$. Therefore, the signal in the lattice data frequently ends considerably before an asymptotic exponential regime is established.

It should be noted that the asymptotic exponential regime is reached more quickly for heavier-than-physical quark masses and smaller momenta $P_z$. On top of that, both heavy quark masses and smaller momenta
exponentially increase the strength of the signal in the lattice data. It is possible that studies with particularly low momentum $P_z$ retain signal until the asymptotic regime is more firmly established, but they are faced with stronger power corrections and perturbative uncertainties in the LaMET formalism.

In the following, we use a dataset with a heavier-than-physical pion mass of 358 MeV. Using this pion mass value as an effective mass for the exponential decay gives a radius $r\sim$ 0.6 fm and with a momentum of 2 GeV an exponential decay in Fourier harmonics $\lambda_r \approx 6$. Despite the improvement in signal and reduction of the decay length, the signal in our dataset still ends shortly after one decay length. The choice of heavier-than-physical pion masses ensures that our conclusions are conservative: these conclusions are likely to be strengthened at physical pion masses.

Particular attention must therefore be devoted to the inverse problem in the transition region between reliable lattice data and asymptotic decay. Although the numerical value of $\lambda$ for this transition region depends on the particular set of quark masses and the hadron momenta, it broadly corresponds to $\lambda \sim 5 - 15$ for studies of phenomenological relevance. The importance of the inverse problem in the closely related short-distance factorization (SDF) framework~\cite{Braun:2007wv,Radyushkin:2017cyf,Ma:2017pxb} is widely acknowledged, and new techniques to handle it are still being explored, but not all are of the same quality. For example,  in closure tests applied to multiple methods of PDF extraction in Ref.~\cite{Karpie:2019eiq}, the reconstructed error consistently underestimated the error of the synthetic data. Simple parametric forms often used to analyze the matrix elements within the SDF framework likewise do not produce generally reliable uncertainty quantification~\cite{Dutrieux:2024rem}.

In this paper, we emphasize that the problem extends to the LaMET framework and make the case for a more comprehensive study of the uncertainty in the transition region than that usually proposed in the literature. The ill-posed nature of the inverse problem in the LaMET framework is only occasionally recognized in the literature (for instance, in \cite{Bhattacharya:2023nmv}), and many publications propose a narrow investigation of its uncertainty. Ref. \cite{Ji:2020brr}, which played an important role in popularizing the use of few-parameter exponential models to extrapolate the missing data, warned that ``to make
the extrapolation under control, the lattice data must exhibit the
exponential decay before the error becomes too big''. The early loss of the signal makes it unlikely that this condition can be fulfilled for many lattice datasets, and the proposed method of fitting the rigid parametric model on various subsets of points gives only a limited account of the uncertainty. The inverse problem of constructing a quasi-PDF from lattice data is therefore ill-posed, not just from a pedantic mathematical point of view, but also from a very practical standpoint, as we will show: 

Even with the assumption of asymptotic exponential decay with physically relevant parameters, the quality of the data in many studies is such that reasonable changes in the treatment of the extrapolation result in significant changes in the uncertainty assessment of the $x$-reconstruction, even in the moderate $x$-range.

Therefore, it appears to us that the question of the $x$-reconstruction in both LaMET and SDF within the precision of currently available data is a fundamentally similar mathematical problem. There exist many different choices in the analysis of lattice data that have complicated impacts on the uncertainty propagation in both frameworks. These choices include, for example, the precise range in Fourier space, the treatment of renormalization, matching and power corrections, and the inclusion of additional assumptions on the $\lambda$-space behavior, such as exponential decay, or the $x$-space behavior at low and high $x$. However, the statement that the LaMET framework allows a direct calculation of the PDF at a value of $x$ while the SDF only gives access to Mellin moments or indirect inference of the $x$-dependence of the PDF \cite{Ji:2022ezo} seems inaccurate to us, as we demonstrate.

In the next section, we explore the connection between Fourier extrapolation and $x$-reconstruction beyond the rigid few-parameter fit, using a real lattice dataset as a testbed. We hope that it will give a sense of a path towards more realistic uncertainty assessment in both frameworks. Informed by the results in practice, we discuss the role of exponential suppression in the inverse problems for LaMET and SDF. Finally we end with concluding statements.

\section{Numerical exploration of the connection between Fourier and $x$-reconstruction}

\begin{figure}
    \centering
    \includegraphics[width=0.95\linewidth]{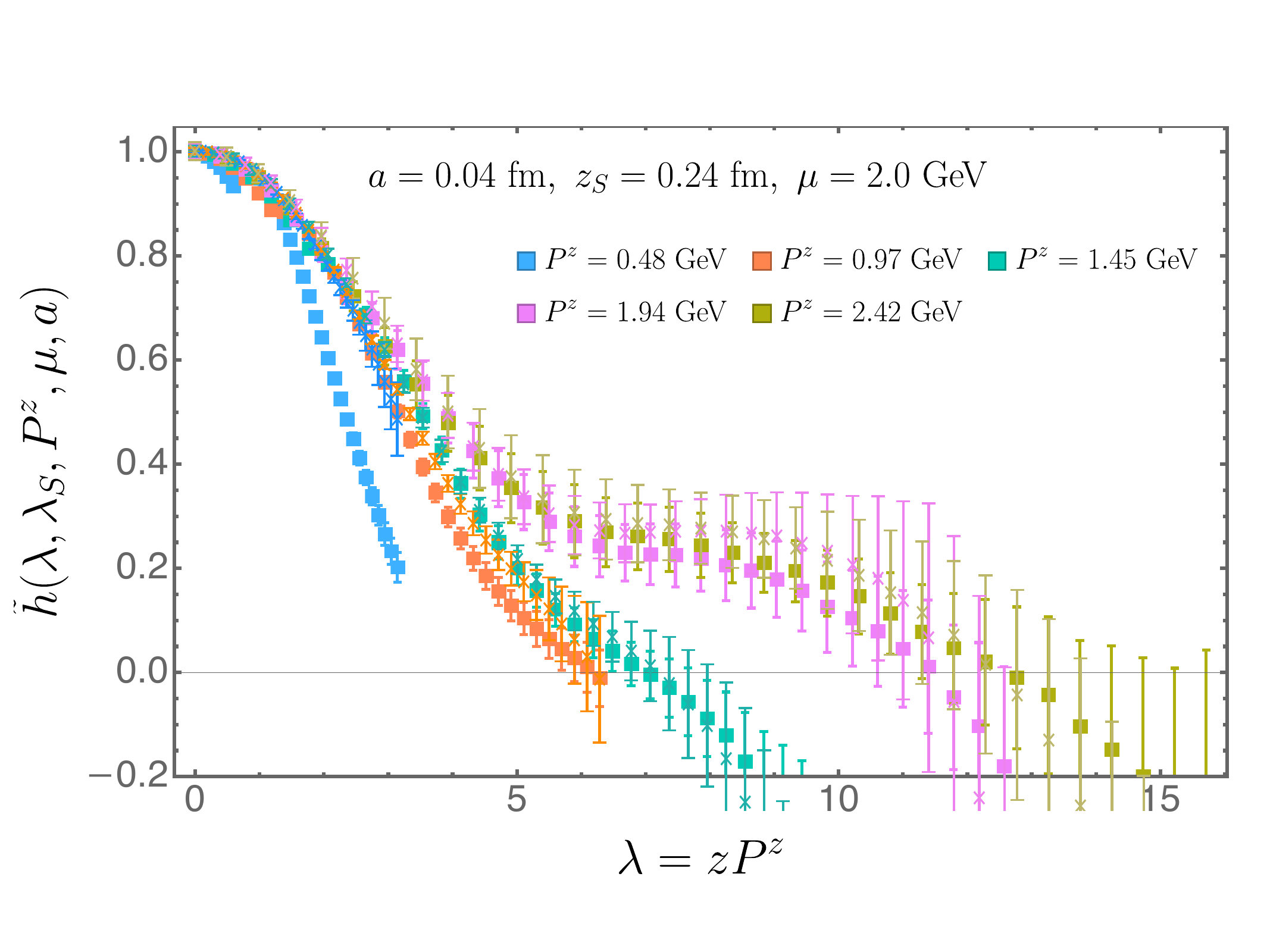}
    \caption{A reproduction of Fig. 1 of Ref.~\cite{Gao:2021dbh} showing the results renormalized in the hybrid scheme (Squares). Overlaid is data in a renormalization group invariant ratio  (Crosses) which are the same as Fig 21 of Ref.~\cite{Gao:2020ito} with larger $z$ extent. Only the lowest two momenta, ignored by subsequent LaMET analysis in Ref~\cite{Gao:2021dbh}, show a statistically resolvable difference in the large $z$ data. It is clear that the two datasets do not have a significantly different behavior at large $z$ for relevant momenta.\label{fig:hybrid_v_ratio}}
\end{figure}

For our numerical exploration, we use matrix elements of the proton unpolarized isovector PDF computed in Ref.~\cite{Egerer:2021ymv} with a single lattice spacing of $0.094$ fm and a pion mass of $358$ MeV. We use a proton momentum $P_z = 2$~GeV and non-local separations $z$ up to 1.13~fm. We use the ratio method to construct matrix elements with a UV-safe limit. Different groups use other strategies to obtain renormalized or UV-safe matrix elements due to their different perspective on the best use of the lattice matrix elements. The ratio method is anticipated to cancel some of the higher-twist effects \cite{Braun:2024snf}, preserving the dominance of the leading-twist contribution up to larger values of $z$, as observed using a spectator diquark model in Fig. 19 of Ref.~\cite{Musch:2010ka}.  An example of an alternative strategy, the hybrid scheme \cite{Ji:2020brr}, aims at a minimal removal of the linear divergence, to preserve the fast decay of the renormalized matrix element at large $z$. However, the current range and precision of the lattice data in most studies are insufficient to distinguish a clear asymptotic trend in renormalized matrix elements, regardless of the renormalization scheme. It should be noted that Ref.~\cite{Chen:2025cxr} criticizes the ratio method for use in LaMET. Fig.~\ref{fig:hybrid_v_ratio} shows the comparison of data generated by some of the authors of Ref.~\cite{Chen:2025cxr} in a previous publication~\cite{Gao:2021dbh}. All the large $P_z$ results used for LaMET in Ref~\cite{Gao:2021dbh} are clearly in statistical agreement with their ratio counterpart. The data generated by some of the authors of Ref.~\cite{Chen:2025cxr} disproves their claim that the ratio ``completely distorts the non-perturbative physics'' for the currently available range of $z$ and statistical precision. While the convergence is in theory asymptotically different, the large noise at large $z$ makes it irrelevant for the present purpose of studying the inverse problems. Our own ratio matrix elements do not seem to decay noticeably less quickly than data with hybrid renormalization, so we believe that our dataset provides a realistic testbed for the sake of this numerical exploration.

\subsection{Reconstruction without explicit constraints\\on the tail asymptotic decay}

Reliable lattice data frequently end before or close to one decay length in Fourier harmonics for phenomenologically relevant analyses. Therefore, the use of a rigid exponential parametric model to complete the missing harmonics introduces a model dependence that needs careful assessment. Before we study reconstructions with a controlled asymptotic behavior in Fourier space, it is useful, for the sake of comparison, to study what happens if no asymptotic criterion is enforced. Besides simple parametric fits of the full $x$-dependence, whose model dependence is acknowledged, several non-parametric methods have been used in the literature, such as the Maximum Entropy Method~\cite{mem_ref}, Bayesian Reconstruction~\cite{Burnier:2013nla, Liang:2019frk}, and Neural Network analyses~\cite{Karpie:2019eiq,Cichy:2019ebf,DelDebbio:2020rgv,Khan:2022vot,Chowdhury:2024ymm}. We focus in this section on the Backus-Gilbert (BG) \cite{BG} method because of its regular use, and the Gaussian process regression (GPR), which we consider as a particularly promising avenue. The connection between the two formalisms has been studied recently in \cite{DelDebbio:2024lwm}.

The BG method is a non-parametric strategy used to handle the inverse Laplace transform for the hadronic tensor \cite{Liang:2017mye, Liang:2019frk} or the inverse Fourier transform of parton distributions, \textit{e.g.}~in~\cite{Karpie:2019eiq,Bhat:2020ktg,Alexandrou:2021bbo, Bhat:2022zrw, Bhattacharya:2023nmv,Bhattacharya:2023jsc}, and has been widely used in other fields. Fundamentally, the method constructs an approximate smeared inverse Fourier operator obtained from the kinematics of the dataset itself. We refer to Section 3.1 of \cite{Karpie:2019eiq} for a compact presentation of the traditional use of the method. The inverse Fourier operator is only approximate, so Fourier transforming the BG $x$-reconstruction can lead to significant discrepancies with the lattice dataset: the traditional use of the BG method in the literature for parton distributions is biased. With a very precise dataset, the bias can be of many standard deviations. This was highlighted in Fig.~5 of \cite{Karpie:2019eiq} and is manifest in our study. To correct for this bias, an operation known as preconditioning can be employed. Typically, the data is pre-fitted by a parametric model to find a dataset-specific modification of the Fourier operator that reduces the bias. The impact of preconditioning was highlighted in \cite{Bhat:2020ktg}, where statistically significant differences in the final $x$-reconstruction were reported, depending on the use or not of preconditioning. The use of a parametric model to attempt to reduce the bias of the traditional BG method reduces its model-independent appeal. We note, however, that more sophisticated uses of the BG method, where the smearing is not an output of the procedure but rather defined by the user, may provide a different avenue to control the bias, see \textit{e.g.}~\cite{Hansen:2019idp}.

Beyond the issue of the bias in the reconstruction, the uncertainty estimated by the BG method can behave irregularly. The necessity to regularize ill-conditioned matrix inverses as an intermediate step introduces an arbitrariness in the exact choice and intensity of the regularization, and directly results in differences in the final $x$-reconstruction uncertainty. An irregular behavior of the uncertainty in the BG method, where the uncertainty tightens strongly for some values of $x$ regardless of the preconditioning, was reported in \cite{Bhat:2020ktg}. 
However, it must be noted that the issue of sharp uncertainty variation at certain values of $x$ is common across many reconstruction strategies applied to lattice data, ranging from neural networks (\textit{e.g.} \cite{Gao:2022uhg}) to parametric fits \cite{Dutrieux:2024rem}. In contrast, GPR provides a more principled framework for managing such artifacts and offers improved control over uncertainty estimates.

\begin{figure}
    \centering
    \includegraphics[width=0.95\linewidth]{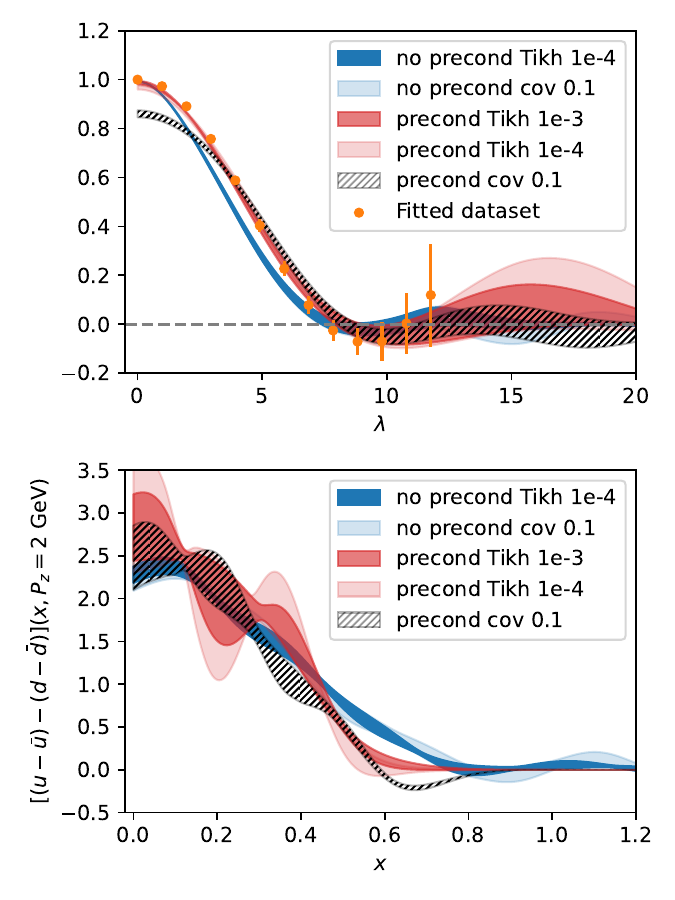}
    \caption{(top) Various BG reconstructions of the same Fourier dataset, either using no-preconditioning (dark and light blue) or with a preconditioning given by the best fit of the data to $N x^\alpha (1-x)^\beta \theta(1-x)\theta(x)$. The use of preconditioning results in very significant differences for $x \approx 0.6$. We also study different regularizations of matrix inverses, either by adding a small multiple of the identity matrix (labelled as ``Tikh'' followed by the used multiplicatory factor) or 0.1 times the data covariance (labelled as ``cov 0.1''). -- (bottom) the corresponding $x$-reconstructions.}
    \label{fig:BG}
\end{figure}

These difficulties with the BG method appear clearly with our dataset in Fig.~\ref{fig:BG}. In the absence of preconditioning, the bias is evident: the proposed $x$-reconstruction is clearly in strong tension with the lattice dataset. For a regularized lattice data covariance \footnote{Since the value at $\lambda = 0$ is exactly 1 by construction of the dataset, the standard $\chi^2$ is infinite without regularization. Therefore, we modify the data covariance to attribute an uncorrelated standard deviation of $10^{-3}$ to that point. In contrast, applying a simple SVD cut would effectively exclude the 
$\lambda=0$ point from contributing to the $\chi^2$.}, the correlated $\chi^2$ of the reconstruction divided by the number of points in the dataset is over $10^3$. With a preconditioning performed thanks to a fit $N x^\alpha (1-x)^\beta \theta(1-x) \theta(x)$ where $\theta$ is the Heaviside step-function, the situation depends on the regularization of the inverse matrix in the intermediate step of the reconstruction. When adding a small multiple of the identity matrix (a form of Tikhonov regularization), the agreement with the dataset increases in appearance, but the regularized correlated $\chi^2$ remains at similar levels. On the other hand, using the data covariance as a regulator produces extreme discrepancies at small $\lambda$, which worsen even more the $\chi^2$. Further explorations on different datasets can be found in Figs. 7.35-7.43 of~\cite{Karpie:2019doj}. We also observe an irregular reduction of the uncertainty in some of our reconstructions.

Although a more in-depth study of the BG method could probably solve some of the issues of the traditional implementation, another non-parametric strategy has been proposed with a more intuitive control over the quality of the reproduction of the lattice data and the properties of the uncertainty in $x$-space. The GPR  with a kernel in $x$-space was explored recently in \cite{Candido:2024hjt, Dutrieux:2024rem}. The general principles are straightforward: a fine grid is chosen in $x$ offering a very large number of degrees of freedom for the reconstruction, here 100 values spanning regularly $x \in [0, 2]$ \footnote{We only consider positive values of $x$ since we study separately the real and imaginary parts of the matrix elements, which correspond to definite parities in $x$.}. To select sensible reconstructions, a Bayesian prior on the covariance of the points in $x$ is added. One can choose a simple Radial-Basis Function (RBF) kernel to correlate the values in $x$, characterized by a variance $\sigma^2$ and a correlation length $l$:
\begin{equation}
K_{\rm RBF}(x, x';\sigma^2,l^2) = \sigma^2 \exp\left(-\frac{(x-x')^2}{2l^2}\right)\,. \label{eq:RBF}
\end{equation}
Of interest to the physics that we are studying may be a logarithmic RBF kernel defined by:
\begin{equation}
K_{\rm log RBF}(x, x';\sigma^2,l^2) = \sigma^2 \exp\left(-\frac{(\log(x)-\log(x'))^2}{2l^2}\right)\,, \label{eq:logRBF}
\end{equation}
which allows for more flexibility at small $x$. Then, a simple correlated least-squares minimization is performed, including both the prior covariance kernel and mean function in $x$ and the dataset of Fourier harmonics. Technical details are laid out in Ref.~\cite{Dutrieux:2024rem}.

\begin{figure}
    \centering
    \includegraphics[width=0.95\linewidth]{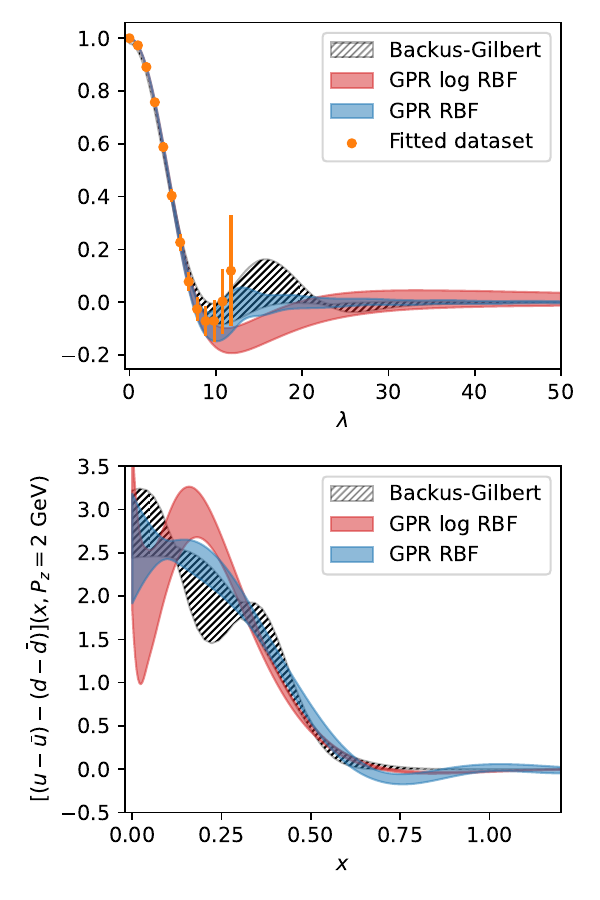}
    \caption{(top) Two GPR reconstructions/extrapolations of the same Fourier dataset using two different covariance kernels in $x$-space corresponding to Eq.~\eqref{eq:RBF} in blue and Eq.~\eqref{eq:logRBF} in red -- (bottom) the corresponding $x$-reconstructions. The Backus-Gilbert solution with preconditioning and a Tikhonov regularization of $10^{-3}$ is shown for comparison.}
    \label{fig:fig1}
\end{figure}

Fig.~\ref{fig:fig1} displays the result of two $x$-reconstructions with the two different kernels with fixed kernel parameters so that the prior in $x$ allows an excellent reproduction of the fitted dataset \footnote{For the RBF kernel, we use $\sigma = 10$, $l = 0.4$ and a prior mean of $f(x) = 10$. For the logarithmical RBF, we use $\sigma = 1$, $l = 0.7$ and a prior mean $f(x) = x^{-0.15}$.}. The regularized correlated $\chi^2$ divided by the number of points is now below 1 in both cases. Strikingly, the RBF extrapolation (blue band) is generally narrower in Fourier space than the logarithmic RBF (red band), especially in the asymptotically large Fourier harmonics, but gives rise to a larger uncertainty for $x > 0.2$. The uncertainty for $x$ close to 1 of the RBF reconstruction is seven times larger than that of the logarithmic RBF reconstruction. The fact that a much narrower tail in $\lambda$ does not translate into a smaller uncertainty for large $x$ indicates that the large $\lambda$ region plays only a minor role in the large $x$ reconstruction, a statement that will become clearer with the next examples. 
Since the small $\lambda$ region is tightly constrained by the fitted dataset and thus nearly identical across both reconstructions, it is crucial to focus on the intermediate range, here $\lambda \in [8, 15]$, where the lattice data is less precise and multiple plausible tail extrapolations remain compatible with the data uncertainties. 

The comparison with the most sensible BG reconstruction that we obtained at the previous step (black hatches in Fig.~\ref{fig:fig1}) shows that the GPR method seems less sensitive to artifacts such as uncertainty tightening, whether in $x$ or Fourier space. This is not surprising since the GPR precisely enforces a requirement of correlation length, reducing the creation of uncertainty nodes which result from strong local anti-correlations.

Given the challenge of reconstruction without explicit constraints on the asymptotic decay tail, it is natural to test whether the addition of explicit constraints changes significantly the severity of the inverse problem, as is sometimes claimed in the LaMET literature~\cite{Gao:2021dbh,Ji:2022ezo}.

\subsection{Biases of truncated models for data extrapolation}

\begin{figure}
    \centering
    \includegraphics[width=0.95\linewidth]{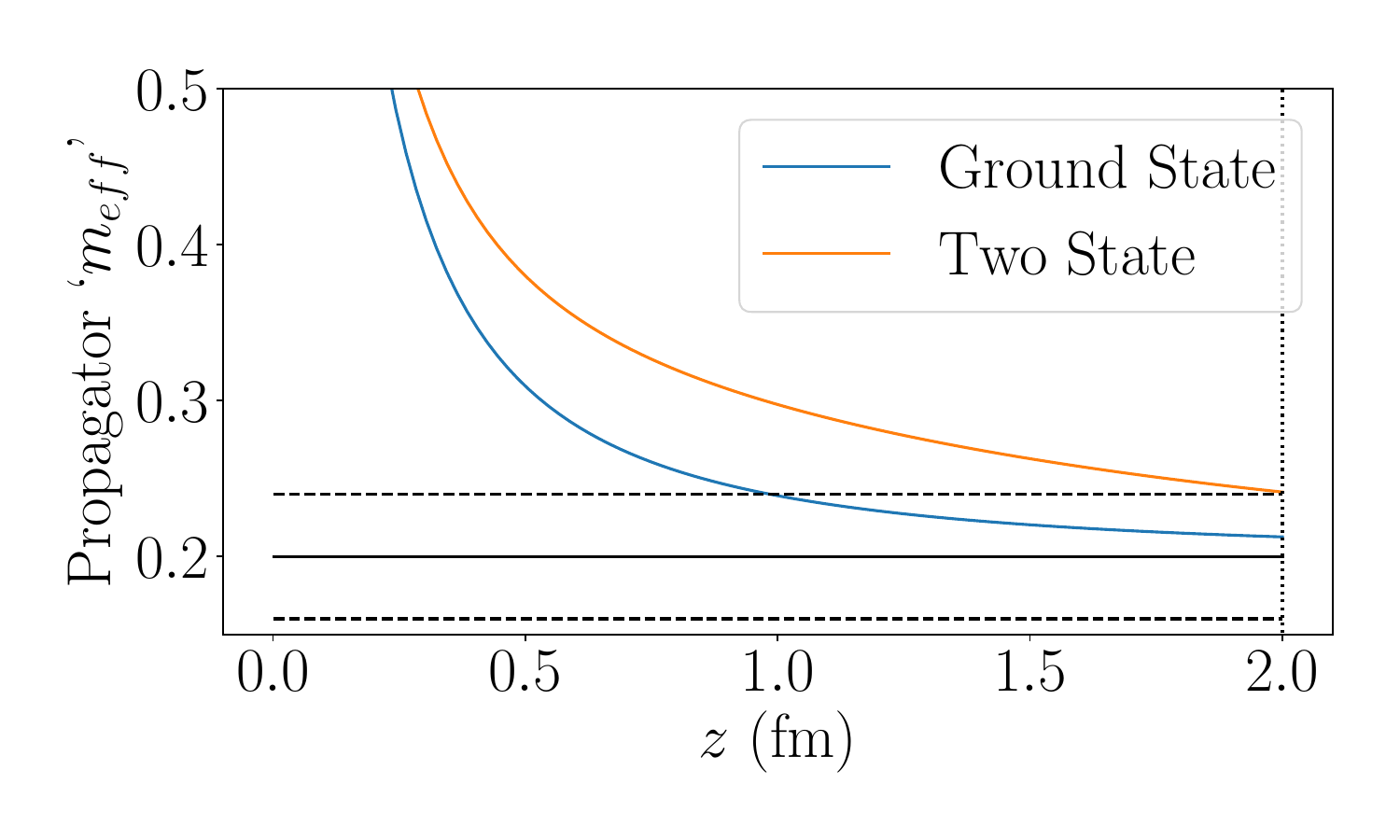}
    \caption{The effective mass of the large $z$ model given a single state (blue) and two states (orange). The solid line is the input mass, and the dashed line represents a 20\% overestimation. \label{fig:prop_meff}}
\end{figure}

Before describing them, we wish to give a warning about the nature of this exponential dependence and how an inappropriate use where the approximations are invalid could lead to model bias and underestimated error. Lattice QCD practitioners handle truncated asymptotic models for correlation functions in the form of a series of exponentials. Inferring matrix elements and masses from this series is an inverse problem which requires a scientist-made choice of truncation, causing a bias known as ``excited state contamination''. If the asymptotic approximation is poor for a given dataset due to too strong a truncation, the bias can be significant and errors underestimated. In the case of this matrix element, the asymptotic $z$ behavior comes from a series of mesons propagating in the space-like distance. The approximation of Refs.~\cite{Gao:2021dbh, Chen:2025cxr} is to drop all but the lightest meson's contribution and further take only the leading asymptotic form of the Bessel function in the meson propagator 
\begin{equation}
        G(z,m) = \frac{-i m}{4\pi^2 |z|} K_1(m|z|) 
        \xrightarrow[z\to\infty]{} A\frac{e^{-m|z|}}{z^{3/2}} \,.\label{eq:bessel approx}
\end{equation} 
Fig.~\ref{fig:prop_meff} shows an effective mass based upon this approximation 
\begin{equation}
    m_{\rm eff}  = -\frac{d}{dz} \log(z^{3/2} G(z,m))
\end{equation}
which should plateau to the mass $m=0.2$ GeV at large $z$ when the approximation is valid. The effective mass for a single meson is systematically overestimated in the region, $z\in[0.75,1.25]$ fm, where real lattice data exists, and this model is applied. The situation worsens when one includes the subdominant second state with twice the mass of the lower state. This choice of mass was made for convenience because a two-particle heavy meson and pion state, or possibly some low excitation, may have such an energy gap near $0.2$ GeV. The same overlap is used since the point sources, like those in the ``heavy quark'' approximation used for the Wilson Line, tend to have similar overlap to all states. 

It was argued in Ref.~\cite{Dutrieux:2025axb} that overestimation of the mass can lead to dramatic underestimation of extrapolation error in Ref.~\cite{Gao:2023ktu} and the inflexibility of a model that can't change sign lead to biased results inconsistent with the trend of the data in Ref.~\cite{Gao:2021dbh}. 

The methods of enforcing an exponential dependence through Bayesian priors proposed below would also be capable of similar underestimation of error or model bias, as these are generic issues of inverse problems. The advantage is GPR allows the single meson approximation to have the flexibility to correctly estimate the subdominant $z$ dependence removed in the approximation of Eq.~\ref{eq:bessel approx}. More complex priors than those below can be generated as the subdominant behavior is better understood to create a systematically improvable procedure, just as the asymptotic approximation of Ref.~\cite{Gao:2021dbh,Chen:2025cxr} while maintaining the advantage of flexibility to have less model bias than the approximate models. Most importantly, the method for including asymptotic exponential convergence provided in our work can be applied in a regime far away from the precision data if that is what is required. Asymptotic modeling, as Ref.~\cite{Gao:2021dbh,Chen:2025cxr} advocate for, can only be applied where the data exist.

\subsection{Exponential and Gaussian tail reconstructions}

Gaussian processes provide a convenient way to add prior information into a Bayesian inference. 
Just as a constraint can be added in $x$ space in terms of the prior mean and covariance kernel, an additional constraint can be added to induce decay in $(\lambda,z)$. While described as a pair to indicate the physical origin of the choices, in practice, with the fixed momentum parallel to the separation, $\lambda$ and $z$ are not independent. In this section, we explore three different types of decay.

For the first exponential strategy, in addition to the previous prior in $x$-space, we add a new GPR prior which is a Gaussian with zero mean and a covariance kernel characterized by a variance $\Sigma^2$, a correlation length $L$, and an exponential decay rate $r$:
\begin{eqnarray}
    &\hspace{-100pt}k_{\rm exp}((\lambda,z),(\lambda',z'); \Sigma^2 , L^2, r) \nonumber\\
    &= \Sigma^2\exp\bigg[-\displaystyle\frac{(\lambda-\lambda')^2}{2L^2}  -  \frac{|z|+|z'|}{r} \bigg]
\end{eqnarray}
to induce a linear exponential decay. Depending on the quark masses, we expect a radius of the order of 1 fm or less to be relevant, corresponding to a decay length in Fourier harmonics $\lambda_r = rP_z$ of up to 10 for $P_z = 2$ GeV. Therefore, the intermediate $\lambda \in [8, 15]$ region that we have identified earlier is still in the transition region towards asymptotic behavior. We vary the radius in our numerical applications, using 0.6 fm, which corresponds to our pion mass, up to 1 fm, which is more appropriate for physical quark masses. \footnote{For the RBF kernel in $x$-space, we use $\sigma = 50$, $l = 0.4$ and prior mean of $f(x)= 3$. For the logarithmical RBF kernel in $x$-space, we use $\sigma = 4$, $l = 0.7$ and prior mean of $f(x)=x^{-0.15}$. For an exponential decay in $\lambda$ of $r = 1.0$ fm, we use a kernel in $\lambda$-space with $\Sigma = 0.5$ and $L = 5$. For an exponential decay of $r = 0.6$ fm, $\Sigma = 1$ so that the prior decay uncertainty coincides at the beginning of the extrapolation region.} We observe little impact of this radius on the actual uncertainty in $x$ in the relevant range $x > 0.2$, as expected from our observations in the previous section that the uncertainty at large $\lambda$ plays a minor role.

\begin{figure}[h!]
    \centering
    \includegraphics[width=0.95\linewidth]{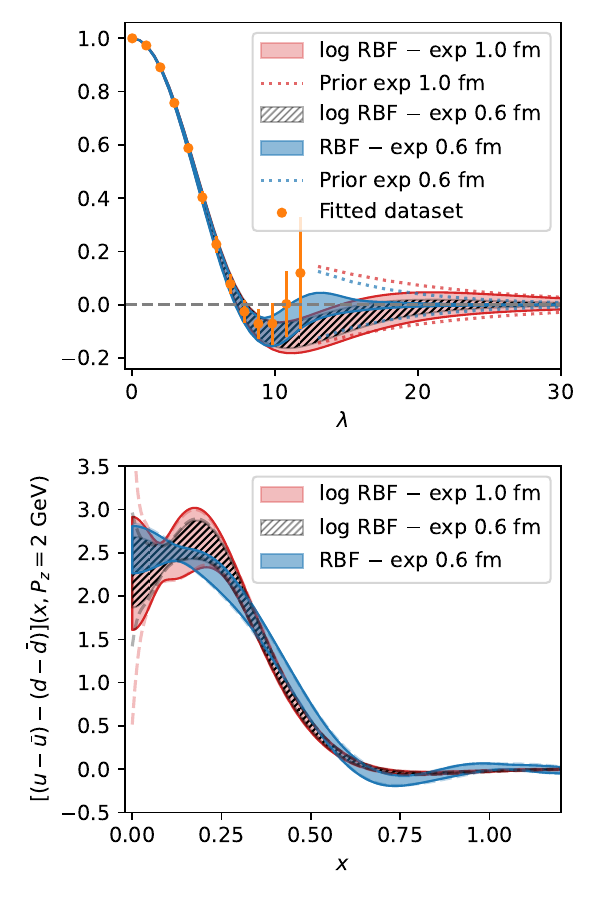}
    \includegraphics[width=0.95\linewidth]{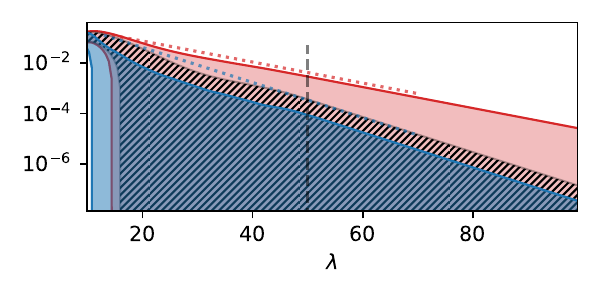}
    \caption{(top) Reconstructions/extrapolations of the same Fourier dataset enforcing an exponential decay with $z$ of length 1.0 fm or 0.6 fm. The decay prior is represented by the dotted lines. -- (middle) the corresponding $x$-reconstructions. The dotted lines represent the GPR reconstruction if an exact exponential tail is not used beyond $\lambda = 50$. -- (bottom) Fourier tail extrapolations displayed in logarithmic scale and with an overall minus sign (since the extrapolation is mainly negative). The black dotted line at $\lambda = 50$ represents the threshold where an exact exponential tail replaces the Gaussian process extrapolation. The colored dotted lines represent the prior enforcing the exponential decay as in the top plot.}
    \label{fig:fig2}
\end{figure}

The prior covariance in $\lambda$ is enforced starting from the last datapoint in the dataset and up to a large asymptotic value, that is here $\lambda \in [13, 70]$. When $\lambda$ becomes larger than 50, we even replace the GPR in Fourier space by an exact exponential decay to make sure that we capture the exact desired asymptotic behavior. We observe that this procedure only has a minute effect on the reconstruction below $x < 0.1$, as depicted by the dotted lines in the $x$-reconstruction of Fig.~\ref{fig:fig2}.

Beyond the asymptotic exponential decay of the matrix element, a suppression at large $z$ can arise from the finite size of the proton and be implemented by the alternative prior kernel:
\begin{eqnarray}
    &\hspace{-100pt}k_{\rm gauss}((\lambda,z),(\lambda',z'); \Sigma^2, L^2, r) \nonumber\\
    &=\Sigma^2 \exp\bigg[-\displaystyle\frac{(\lambda-\lambda')^2}{2L^2}  + \frac{z^2+z'^2}{2r^2}  \bigg]
\end{eqnarray}
with spacelike $z^2,z'^2<0$ that force a Gaussian decay at a rate dictated by the proton radius, $r$. Here, we choose $r = 0.8$ fm for a practical application.

\begin{figure}
    \centering
    \includegraphics[width=0.95\linewidth]{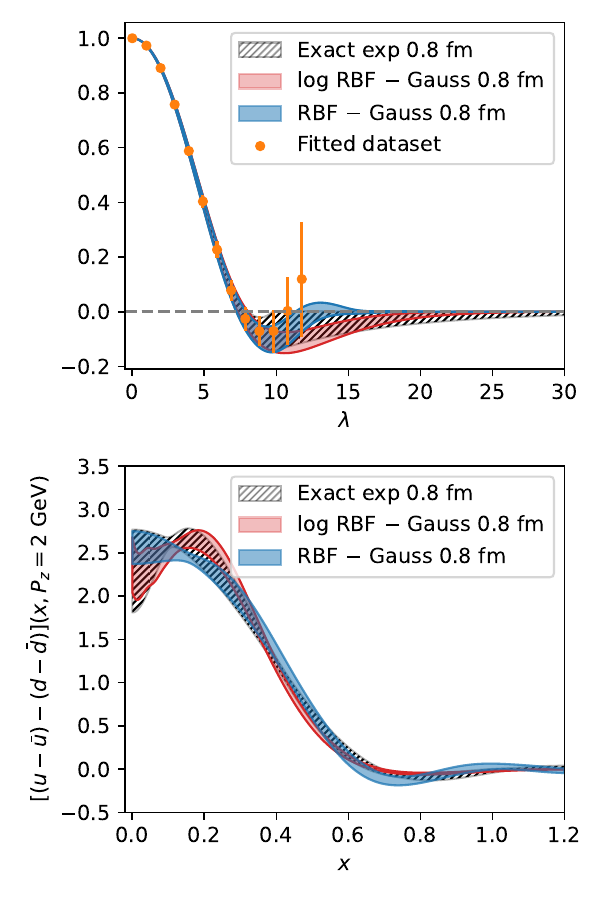}
    \caption{(top) Reconstructions / extrapolations of the same Fourier dataset, either replacing directly the data by an exact exponential tail or using a Gaussian process reconstruction with a Gaussian decay with radius 0.8 fm -- (bottom) the corresponding $x$-reconstructions.}
    \label{fig:fig4}
\end{figure}

Finally, we consider a third strategy of rigid parametric extrapolation, where we interpolate the lattice data up to $z = 0.94$ fm, or $\lambda \approx 10$, where the data have decayed to nearly 0 and the uncertainty starts to sharply increase. We extrapolate to the large harmonics with the simple model $A \exp(- 0.12 \lambda)$ where $A$ is fitted for continuity. Since we do not believe that our data allows us to reliably fit the coefficient of an exponential decay, we fix the coefficient $0.12$ to correspond to an exponential decay in $z$ of $r = 0.8$ fm for $P_z = 2$ GeV.

 The results are displayed in Figs.~\ref{fig:fig2} and \ref{fig:fig4}. Fig.~\ref{fig:fig2} shows that, in the range of interest ($x > 0.2$) for $P_z = 2$ GeV, the effect of the choice of kernel in $x$ matters significantly more than the choice of asymptotic exponential decay. In fact, the $x$-reconstructions share many similarities with those of Fig.~\ref{fig:fig1}, which used the same kernels in $x$, but for which no explicit constraint was enforced on the asymptotic tail behavior at all. In Fig.~\ref{fig:fig4}, the GPR reconstructions with Gaussian tails exhibit once more similar characteristics to those with or without an exponential tail constraint, provided they use the same kernel in $x$-space. On the other hand, the tail extrapolation obtained by the simple parametric exponential model beyond $\lambda = 10$ presents a sufficiently different reconstruction to have some regions in $x$ with 1-$\sigma$ tension.

\begin{figure}
    \centering
    \includegraphics[width=0.95\linewidth]{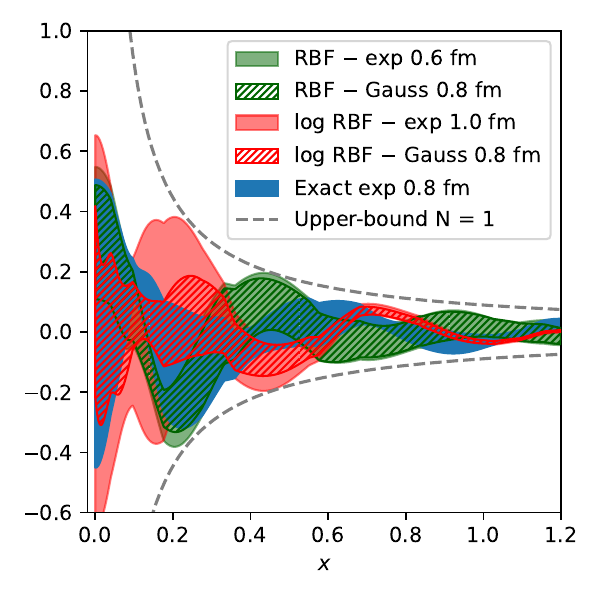}
    \caption{Various $x$ reconstructions with controlled asymptotic tail behavior discussed in this document, where all curves have been subtracted by the same mean function $f(x)$ -- the mean of the envelope of all curves -- to better gauge the absolute size of the uncertainty and the compatibility of the various reconstructions. The dotted lines represents the ``upper-bound'' on uncertainty derived in Ref.~\cite{Gao:2021dbh}, namely $4 N_x |h(\lambda_L)| / (\pi x)$ where we used $N_x = 1$, $\lambda_L = 10$ and $|h(\lambda_L)| = 0.07$. }
    \label{fig:fig5}
\end{figure}

A better way to appreciate that, for $x > 0.2$, the choice of methodology (using a given $x$-space kernel for a GPR or replacing the tail by a parametric model) matters much more to the central value than the asymptotic tail decay rate is the comparison plot of Fig.~\ref{fig:fig5}, where the average has been subtracted for legibility. Assuming an exponential decay, a Gaussian decay, or playing with the radius within a reasonable phenomenological range leads to very little change to the central value. On the other hand, the true effect of a different kernel in $x$-space, as can be assessed in Figs.~\ref{fig:fig1}, \ref{fig:fig2}, and \ref{fig:fig4}, is to produce significantly different behaviors in the intermediate $\lambda \in [8, 15]$ region. Using any single method would result in an unreliable representation of the uncertainty, and small parametric variations, such as using different effective masses, do not appear sufficient to build a reliable uncertainty quantification. We believe that the large difference in the uncertainty of the reconstructions displayed in Fig.~\ref{fig:fig5} demonstrates that there is indeed an inverse problem that must be handled seriously. We would like to specifically note that the nature of the exponential does have a strong impact on the variance of the result. Incorrect application of the exponential dependence, such as overestimating the mass by fitting in the subasymptotic regime, could lead to incorrect uncertainty quantification.

\begin{figure}
    \centering
    \includegraphics[width=0.95\linewidth]{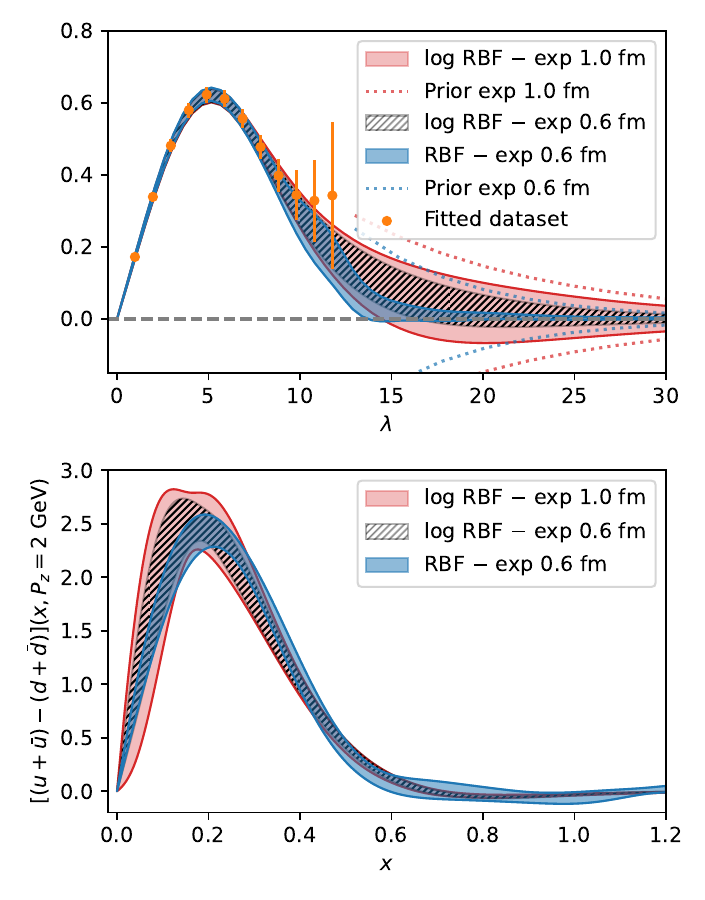}
    \caption{\label{fig:imag_fit} (top) Reconstructions/extrapolations of the imaginary part of the Fourier dataset enforcing an exponential decay with $z$ of length 1.0 fm or 0.6 fm. The decay prior, represented by the dotted lines, has a standard deviation twice as large as that used for the real part in Fig.~\ref{fig:fig2}. -- (bottom) the corresponding $x$-reconstructions. Notice that the exponential suppression of the large Fourier components for an odd function of $x$ makes the function go smoothly to 0 at $x = 0$.}
\end{figure}

The real component of the matrix element, studied above, is significantly more convergent than the imaginary component, as seen in Fig.~\ref{fig:imag_fit}. It was suggested in Ref.~\cite{Chen:2025cxr} that LaMET should only be performed when data have satisfied a sufficient level of convergence, which is hardly ever obtained to date in lattice QCD calculations of the imaginary component of the matrix element. Soon afterwards, in Ref.~\cite{Xiong:2025obq}, some of the authors of Ref.~\cite{Chen:2025cxr} suggested that methods which can handle imprecise data are actually useful, even proposing a special case of GPR previously provided in this manuscript. The imaginary component is a clear example of such a setting, as neglecting half of one's dataset would be a large waste of computing resources and an unfortunate physical choice.

In Fig.~\ref{fig:imag_fit} we show the results of the Gaussian Process with exponentially decaying prior distributions to demonstrate how the physical decay behavior could be applied with highly non-convergent data. To account for the larger uncertainty in the data, we have doubled the size of the prior, enforcing the exponential decay at large $\lambda$ while keeping every other parameter unchanged compared to the real part. Once again, we observe very little importance of the exact rate of exponential decay, while the choice of regularization in $x$-space generates large discrepancies in the large $x$ region where LaMET is meant to be most precise. 

\section{$x$-reconstruction in LaMET and SDF}

The quasi-PDF definition uses matrix elements with a large value of $z$ that SDF discards due to  off light cone contamination and the breakdown of perturbation theory. While this reduces the severity of the ill-posedness in the Fourier inverse problem within LaMET, the range of Fourier harmonics accessible with sufficient signal quality in current studies remains too limited to disregard the problem, even at large values of $x$. Since the asymptotic behavior of Fourier harmonics has a minimal effect on the reconstruction at large $x$, the mathematical problem that both frameworks face is of a similar nature.

An interesting development towards uncertainty quantification is the derivation of a ``rigorous upper-bound'' on the uncertainty of $x$-reconstruction for the case of an exponential decay in Appendix B of Ref.~\cite{Gao:2021dbh}. This ``upper-bound'' depends on a number $N_x$ of periods of the Fourier weight $\exp(i x \lambda)$ beyond which the authors consider that the oscillating weight effectively suppresses contributions from the large Fourier components for a given value of $x$. They recommend using $N_x = 1$ for a phenomenologically relevant exponential decay of the Fourier components. It appears that any single extraction we have studied in Fig.~\ref{fig:fig5} produces uncertainties significantly below that of this ``upper-bound'' for $N_x = 1$, but the general envelope of all our reconstructions follows fairly well this rule-of-thumb in the region of $x$ of interest. Unlike how Ref.~\cite{Chen:2025cxr} seems to have understood our work, we have never pretended that uncertainty quantification is impossible due to the inverse problem. Rather, we have pointed out the necessity to acknowledge the existence of the inverse problem and stressed that rigid parametric extrapolation, or the BG method, does not faithfully depict that uncertainty when lattice data is limited. We have additionally underlined that the importance attributed to the precise asymptotic decay in LaMET does not appear to be grounded in concrete evidence, as enforcing or not this decay does not change crucially the picture of uncertainty propagation.

This excessive importance devoted to the asymptotic behavior leads to a surprising uncertainty assessment by the authors of Ref.~\cite{Gao:2021dbh} when it comes to evaluating the ``upper-bound'' on uncertainty in the situation of the SDF framework. Having concluded that the quasi-PDF could have an absolute uncertainty of ``less than 0.15 at $x = 0.5$'' by assuming a specific exponential convergence to zero, the authors apply their reasoning to a power-decaying tail and conclude that ``the uncertainty in the FT is of orders of magnitude larger than that of extrapolation with exponential decay.'' This claim was made despite the fact that earlier published studies~\cite{Karpie:2019eiq,Alexandrou:2020tqq} of non-parametric methods without exponential assumptions show in closure tests to fall within their stated bound with $N_x=1$ or 2 in some cases.

It seems unlikely that the absolute uncertainty of any sensible reconstruction for $x > 0.2$ in SDF, where a power-decaying tail is an appropriate assumption, could be orders of magnitude larger than 0.15. It is certainly true that a power-decaying tail will become orders of magnitude larger than an exponentially decaying one for large enough $\lambda$. But as we have stressed before, the asymptotic behavior of the tail represents only a minor contribution to the Fourier transform uncertainty at large $x$. A more realistic bound in Ref.~\cite{Gao:2021dbh} can be obtained by considering that (at least ordinary) PDFs are known to be smooth functions of $x$. The contribution of Fourier harmonics that are much larger than the inverse of the length-scale of fluctuations of the PDF at a certain value of $x$ is then physically suppressed \footnote{In fact, if one uses the criterion proposed in Ref.~\cite{Gao:2021dbh} that Fourier components at large $\lambda$ do not matter to the reconstruction at a given value of $x$ if the derivative of the Fourier transform of the uncertainty is much smaller than $x$ (with their notations, $|\delta \tilde{h}'(\lambda)| \ll x$), one finds a much more sensible depiction of uncertainty propagation in the framework of short-distance factorization. Using their example Eq. (B22) for a non-exponential algebraic decay of the large Fourier components as $\lambda^{-1/2}$, corresponding to a phenomenological PDF which diverges as $x^{-1/2}$ at small $x$, the asymptotic error could be approximated by $|\delta \tilde h(\lambda)| \lesssim \tilde h(\lambda) \approx 0.1 (10 / \lambda)^{1/2}$. Already if 
$\lambda = 10$, the derivative of this bound equals 0.005, so well below any value of $x$ of phenomenological relevance. Unlike what Ref.~\cite{Gao:2021dbh} concludes, their own reasoning would give that short-distance factorization has a controlled asymptotic expansion, resolving the contradiction between their statements and the empirical evidence they provide themselves that the exponential decay does not matter crucially. But the reasoning is also fragile because it relies on asymptotic forms in a regime where they are not established and very approximate bounds: we find that our more in-depth study shows sensitivity beyond $\lambda = 10$ because the approximation by a simple asymptotic model is inaccurate.}.

This perspective offers a useful way to examine the claim that LaMET enables the direct computation of PDFs at fixed values of $x$~\cite{Ji:2022ezo, Ji:2024oka}. Suppose the light cone PDF exhibits a sharp feature at some specific value of $x$. In that case, accurately capturing this behavior would require reliable information in the Fourier domain at scales corresponding to the inverse of the feature's spatial width. However, the quasi-PDF is exponentially suppressed in Fourier modes beyond a cutoff scale roughly proportional to $P_z / \Lambda_{QCD}$. This suppression is rooted in the spacelike behavior of the correlation function at large distance, which is not relevant to the  light cone PDF itself. As a result, LaMET reconstructions are effectively filtered through a smearing kernel with a width on the order of $\Lambda_{QCD} / P_z$, limiting the sensitivity to rapid variations in the light cone PDF. There is no reconstruction of the light cone PDF at a well-defined value of $x$ possible in LaMET. Reliable results can only be obtained through the assumption of smoothness of the target function, a limitation common to any other framework of lattice calculation of parton distributions. In other words, while improved lattice data that extend to higher Fourier harmonics may largely resolve the inverse problem for quasi-PDFs within the relevant $x$-range, reconstructing the true $x$-dependence of the light cone PDF remains obstructed. The practical impact of this complication depends on the exact shape of the distribution in question, how the exact behavior of the most rapidly changing regions, and the ranges of $x$ in which those regions occur. This obstruction arises from missing information at large $z$, a limitation analogous to that encountered in the SDF approach. We note that the notion that the exponential fall-off at large $z$ of the quasi-PDF is an obstruction to the reconstruction of the light-cone PDF has been presented in Ref.~\cite{Ji:2020brr} before.

\section{Conclusions}

The various procedures described above are just a few ways to enforce a desired asymptotic decay among many other kernels or procedures that can be tested to thoroughly study the reconstruction dependence. The observation that the asymptotic behavior is of little relevance compared to the transition region for large $x$ reconstruction helps understand the great confidence that has been derived from fixed parametric tail extrapolations in several recent LaMET studies: they generally produce a very minimal exploration of the transition region. Whereas Gaussian processes provide a simple strategy to explore plausible extrapolations more thoroughly than parametric fits, they are not better than the relevance of their priors, and their model dependence still deserves study. Likewise, the non-parametric Backus-Gilbert method routinely used by some groups to perform the $x$-reconstruction is very sensitive to the preconditioning and performs poorly on a limited dataset. Nonetheless, it can be one of several procedures for obtaining reliable uncertainties.

In this work, we have selected only two prior covariance kernels, a single prior mean function, and performed analysis with fixed hyperparameters. These of course are not the only reasonable options. Ref.~\cite{Medrano:2025cmg} studies a wider set of kernels and provides a comparison of fixing and integrating hyperparameters. Generally, the results of reasonable kernels were quite similar. Future study of the exponential priors used in this work could use the same information criteria to study the choice of exponential rate and form.

The uncertainty linked to the inverse problem of Fourier extrapolation is just one of the many challenges to establishing a full uncertainty budget in the computation of parton distributions on a lattice. Few, if any, studies of non-local matrix elements on the lattice demonstrate simultaneous control of the continuum limit, infinite volume limit, physical pion mass limit, control of higher twist contamination, of excited state contamination, and perturbative matching uncertainties. Therefore, the shortcuts that have been used in many exploratory works regarding the question of the Fourier extrapolation should not be seen as particularly concerning. However, until the technology has evolved sufficiently to obtain a good signal beyond $\lambda \approx 15$, we believe that an acknowledgment and a serious assessment of uncertainty propagation in this inverse problem is necessary. It is interesting to note how the errors in $\lambda$ space relate to those in $x$ space. Specifically, these spaces are the Fourier transform of each other. Future studies can analyze the eigen structure of the covariance in $\lambda$ to identify precise and imprecise degrees of freedom and how they manifest themselves in Ioffe time and momentum fraction spaces.

\section*{Acknowledgments}
This project was supported by the U.S.~Department of Energy, Office of Science, Contract \#DE-AC05-06OR23177, under which Jefferson Science Associates, LLC operates Jefferson Lab. This work has benefited from the collaboration enabled by the Quark-Gluon Tomography (QGT) Topical Collaboration, U.S.~DOE Award \mbox{\#DE-SC0023646}. CJM is supported in part by U.S.~DOE ECA \mbox{\#DE-SC0023047} and in part by U.S.~DOE Award \mbox{\#DE-SC0025908}.
This research was funded, in part (HD and SZ), by l’Agence Nationale de la Recherche (ANR), project ANR-23-CE31-0019. AR acknowledges support by U.S.~DOE Grant \mbox{\#DE-FG02-97ER41028}. 
KO was supported in part by U.S.~DOE Grant \mbox{\#DE-FG02-04ER41302} and would like to acknowledge the hospitality of the American Academy in Rome, where he spent part of his sabbatical. 
The research was conducted in part (DR) under the Laboratory-Directed
Research and Development Program at Thomas Jefferson National Accelerator Facility for the U.S. Department of Energy.
Computations for this work were carried out in part on facilities of the USQCD Collaboration, which are funded by the Office of Science of the U.S.~Department of Energy. This work was performed in part using computing facilities at William \& Mary, which were provided by contributions from the National Science Foundation (MRI grant PHY-1626177), and the Commonwealth of Virginia Equipment Trust Fund. In addition, this work used resources at NERSC, a DOE Office of Science User Facility supported by the Office of Science of the U.S. Department of Energy under Contract \#DE-AC02-05CH11231, as well as resources of the Oak Ridge Leadership Computing Facility at the Oak Ridge National Laboratory, which is supported by the Office of Science of the U.S. Department of Energy under Contract No. \mbox{\#DE-AC05-00OR22725}.The authors acknowledge support as well as computing and storage resources by GENCI on Adastra (CINES), Jean-Zay (IDRIS) under project (2020-2024)-A0080511504.
The software codes {\tt Chroma} \cite{Edwards:2004sx}, {\tt QUDA} \cite{Clark:2009wm, Babich:2010mu}, {\tt QPhiX} \cite{QPhiX2}, and {\tt Redstar} \cite{Chen:2023zyy} were used in our work. The authors acknowledge support from the U.S. Department of Energy, Office of Science, Office of Advanced Scientific Computing Research, and Office of Nuclear Physics, Scientific Discovery through Advanced Computing (SciDAC) program, and from the U.S. Department of Energy Exascale Computing Project (ECP). The authors also acknowledge the Texas Advanced Computing Center (TACC) at The University of Texas at Austin for providing HPC resources, like the Frontera computing system~\cite{frontera}, which has contributed to the research results reported within this paper. The authors acknowledge William \& Mary Research Computing for providing computational resources and/or technical support that have contributed to the results reported within this paper.

\bibliography{biblio.bib}

@article{Dutrieux:2024rem,
    author = "Dutrieux, Herv\'e and Karpie, Joseph and Orginos, Kostas and Zafeiropoulos, Savvas",
    title = "{Simple nonparametric reconstruction of parton distributions from limited Fourier information}",
    eprint = "2412.05227",
    archivePrefix = "arXiv",
    primaryClass = "hep-lat",
    reportNumber = "JLAB-THY-24-4242",
    doi = "10.1103/PhysRevD.111.034515",
    journal = "Phys. Rev. D",
    volume = "111",
    number = "3",
    pages = "034515",
    year = "2025"
}

@article{Medrano:2025cmg,
    author = "Medrano, Yamil Cahuana and Dutrieux, Herv{\'e} and Karpie, Joseph and Orginos, Kostas and Zafeiropoulos, Savvas",
    title = "{Gaussian Processes for Inferring Parton Distributions}",
    eprint = "2510.21041",
    archivePrefix = "arXiv",
    primaryClass = "hep-lat",
    reportNumber = "JLAB-THY-25-4579",
    month = "10",
    year = "2025", journal=""
}

@book{doi:10.1137/1.9780898717921,
author = {Tarantola, Albert},
title = {Inverse Problem Theory and Methods for Model Parameter Estimation},
publisher = {Society for Industrial and Applied Mathematics},
year = {2005},
doi = {10.1137/1.9780898717921},
address = {},
edition   = {},
URL = {https://doi.org/10.1137/1.9780898717921}
}

@book{Aster2013, 
title = {Parameter Estimation and Inverse Problems}, 
author = {Richard C. Aster and Brian Borchers and Clifford H. Thurber}, 
edition = {Second Edition}, 
publisher = {Academic Press}, 
year = {2013}, 
isbn = {978-0-12-385048-5}, 
doi = {10.1016/C2009-0-61134-X}, 
url = {https://doi.org/10.1016/C2009-0-61134-X} 
}

@article{Gao:2020ito,
    author = "Gao, Xiang and Jin, Luchang and Kallidonis, Christos and Karthik, Nikhil and Mukherjee, Swagato and Petreczky, Peter and Shugert, Charles and Syritsyn, Sergey and Zhao, Yong",
    title = "{Valence parton distribution of the pion from lattice QCD: Approaching the continuum limit}",
    eprint = "2007.06590",
    archivePrefix = "arXiv",
    primaryClass = "hep-lat",
    doi = "10.1103/PhysRevD.102.094513",
    journal = "Phys. Rev. D",
    volume = "102",
    number = "9",
    pages = "094513",
    year = "2020"
}

@article{Chen:2025cxr,
    author = "Chen, Jiunn-Wei and others",
    title = "{LaMET's Asymptotic Extrapolation vs. Inverse Problem}",
    eprint = "2505.14619",
    archivePrefix = "arXiv",
    primaryClass = "hep-lat",
    month = "5",
    year = "2025", 
    journal = ""
}

@article{Xiong:2025obq,
    author = "Xiong, Ao-Sheng and Hua, Jun and Wei, Ting and Yu, Fu-Sheng and Zhang, Qi-An and Zheng, Yong",
    title = "{Ill-Posedness in Limited Discrete Fourier Inversion and Regularization for Quasi Distributions in LaMET}",
    eprint = "2506.16689",
    archivePrefix = "arXiv",
    primaryClass = "hep-lat",
    month = "6",
    year = "2025", 
journal = ""
}

@article{Ji:2013dva,
    author = "Ji, Xiangdong",
    title = "{Parton Physics on a Euclidean Lattice}",
    eprint = "1305.1539",
    archivePrefix = "arXiv",
    primaryClass = "hep-ph",
    doi = "10.1103/PhysRevLett.110.262002",
    journal = "Phys. Rev. Lett.",
    volume = "110",
    pages = "262002",
    year = "2013"
}

@article{Clark:2009wm,
    author = "Clark, M.A. and Babich, R. and Barros, K. and Brower, R.C. and Rebbi, C.",
    eprint = "0911.3191",
    archivePrefix = "arXiv",
    primaryClass = "hep-lat",
    doi = "10.1016/j.cpc.2010.05.002",
    journal = "Comput. Phys. Commun.",
    volume = "181",
    pages = "1517--1528",
    year = "2010"
}

@article{Edwards:2004sx,
    author = "Edwards, Robert G. and Joo, Balint",
    editor = "Bodwin, Geoffrey T. and Sinclair, D.K. and Eichten, E. and Holmgren, D. and Kronfeld, Andreas S. and Mackenzie, P. and Okamoto, M. and Simone, J. and El-Khadra, Aida X.",
    collaboration = "SciDAC, LHPC, UKQCD",
    title = "{The Chroma software system for lattice QCD}",
    eprint = "hep-lat/0409003",
    archivePrefix = "arXiv",
    reportNumber = "JLAB-THY-04-54",
    doi = "10.1016/j.nuclphysbps.2004.11.254",
    journal = "Nucl. Phys. B Proc. Suppl.",
    volume = "140",
    pages = "832",
    year = "2005"
}

@article{Gao:2023ktu,
    author = "Gao, Xiang and Hanlon, Andrew D. and Mukherjee, Swagato and Petreczky, Peter and Shi, Qi and Syritsyn, Sergey and Zhao, Yong",
    title = "{Transversity PDFs of the proton from lattice QCD with physical quark masses}",
    eprint = "2310.19047",
    archivePrefix = "arXiv",
    primaryClass = "hep-lat",
    doi = "10.1103/PhysRevD.109.054506",
    journal = "Phys. Rev. D",
    volume = "109",
    number = "5",
    pages = "054506",
    year = "2024"
}

@article{Ding:2024saz,
    author = "Ding, Heng-Tong and Gao, Xiang and Mukherjee, Swagato and Petreczky, Peter and Shi, Qi and Syritsyn, Sergey and Zhao, Yong",
    title = "{Three-dimensional imaging of pion using lattice QCD: generalized parton distributions}",
    eprint = "2407.03516",
    archivePrefix = "arXiv",
    primaryClass = "hep-lat",
    doi = "10.1007/JHEP02(2025)056",
    journal = "JHEP",
    volume = "02",
    pages = "056",
    year = "2025"
}

@article{Khan:2022vot,
    author = "Khan, Tanjib and Liu, Tianbo and Sufian, Raza Sabbir",
    title = "{Gluon helicity in the nucleon from lattice QCD and machine learning}",
    eprint = "2211.15587",
    archivePrefix = "arXiv",
    primaryClass = "hep-lat",
    doi = "10.1103/PhysRevD.108.074502",
    journal = "Phys. Rev. D",
    volume = "108",
    number = "7",
    pages = "074502",
    year = "2023"
}

@inproceedings{Babich:2010mu,
    author = "Babich, Ronald and Clark, Michael A. and Joo, Balint",
    title = "{Parallelizing the QUDA Library for Multi-GPU Calculations in Lattice Quantum Chromodynamics}",
    booktitle = "{SC 10 (Supercomputing 2010)}",
    eprint = "1011.0024",
    archivePrefix = "arXiv",
    primaryClass = "hep-lat",
    reportNumber = "JLAB-IT-10-01",
    month = "11",
    year = "2010"
}

@article{Chowdhury:2024ymm,
    author = "Chowdhury, Talal Ahmed and Izubuchi, Taku and Kamruzzaman, Methun and Karthik, Nikhil and Khan, Tanjib and Liu, Tianbo and Paul, Arpon and Schoenleber, Jakob and Sufian, Raza Sabbir",
    title = "{Polarized and unpolarized gluon PDFs: Generative machine learning applications for lattice QCD matrix elements at short distance and large momentum}",
    eprint = "2409.17234",
    archivePrefix = "arXiv",
    primaryClass = "hep-lat",
    doi = "10.1103/PhysRevD.111.074509",
    journal = "Phys. Rev. D",
    volume = "111",
    number = "7",
    pages = "074509",
    year = "2025"
}

@article{DelDebbio:2020rgv,
    author = "Del Debbio, Luigi and Giani, Tommaso and Karpie, Joseph and Orginos, Kostas and Radyushkin, Anatoly and Zafeiropoulos, Savvas",
    title = "{Neural-network analysis of Parton Distribution Functions from Ioffe-time pseudodistributions}",
    eprint = "2010.03996",
    archivePrefix = "arXiv",
    primaryClass = "hep-ph",
    doi = "10.1007/JHEP02(2021)138",
    journal = "JHEP",
    volume = "02",
    pages = "138",
    year = "2021"
}

@article{Hackett:2024xnx,
    author = "Hackett, Daniel C. and Wagman, Michael L.",
    title = "{Lanczos algorithm for lattice QCD matrix elements}",
    eprint = "2407.21777",
    archivePrefix = "arXiv",
    primaryClass = "hep-lat",
    reportNumber = "FERMILAB-PUB-24-0407-T",
    doi = "10.1103/zjzt-rv86",
    journal = "Phys. Rev. D",
    volume = "112",
    number = "5",
    pages = "054506",
    year = "2025"
}

@article{Wagman:2024rid,
    author = "Wagman, Michael L.",
    title = "{Lanczos Algorithm, the Transfer Matrix, and the Signal-to-Noise Problem}",
    eprint = "2406.20009",
    archivePrefix = "arXiv",
    primaryClass = "hep-lat",
    reportNumber = "FERMILAB-PUB-24-0320-T",
    doi = "10.1103/pcvc-734h",
    journal = "Phys. Rev. Lett.",
    volume = "134",
    number = "24",
    pages = "241901",
    year = "2025"
}

@article{Fischer:2020bgv,
    author = "Fischer, Matthias and Kostrzewa, Bartosz and Ostmeyer, Johann and Ottnad, Konstantin and Ueding, Martin and Urbach, Carsten",
    title = "{On the generalised eigenvalue method and its relation to Prony and generalised pencil of function methods}",
    eprint = "2004.10472",
    archivePrefix = "arXiv",
    primaryClass = "hep-lat",
    doi = "10.1140/epja/s10050-020-00205-w",
    journal = "Eur. Phys. J. A",
    volume = "56",
    number = "8",
    pages = "206",
    year = "2020"
}

@article{Aubin:2011zz,
    author = "Aubin, C. and Orginos, K.",
    editor = "Vranas, Pavlos",
    title = "{An improved method for extracting matrix elements from lattice three-point functions}",
    doi = "10.22323/1.139.0148",
    journal = "PoS",
    volume = "LATTICE2011",
    pages = "148",
    year = "2011"
}

@article{Cichy:2019ebf,
    author = "Cichy, Krzysztof and Del Debbio, Luigi and Giani, Tommaso",
    title = "{Parton distributions from lattice data: the nonsinglet case}",
    eprint = "1907.06037",
    archivePrefix = "arXiv",
    primaryClass = "hep-ph",
    doi = "10.1007/JHEP10(2019)137",
    journal = "JHEP",
    volume = "10",
    pages = "137",
    year = "2019"
}

@article{Bhattacharya:2023jsc,
    author = "Bhattacharya, Shohini and others",
    title = "{Generalized parton distributions from lattice QCD with asymmetric momentum transfer: Axial-vector case}",
    eprint = "2310.13114",
    archivePrefix = "arXiv",
    primaryClass = "hep-lat",
    doi = "10.1103/PhysRevD.109.034508",
    journal = "Phys. Rev. D",
    volume = "109",
    number = "3",
    pages = "034508",
    year = "2024"
}

@article{Burnier:2013nla,
    author = "Burnier, Yannis and Rothkopf, Alexander",
    title = "{Bayesian Approach to Spectral Function Reconstruction for Euclidean Quantum Field Theories}",
    eprint = "1307.6106",
    archivePrefix = "arXiv",
    primaryClass = "hep-lat",
    doi = "10.1103/PhysRevLett.111.182003",
    journal = "Phys. Rev. Lett.",
    volume = "111",
    pages = "182003",
    year = "2013"
}

@article{mem_ref,
    author = "Rietsch, E.",
    title = "{The maximum entropy approach to inverse problems - spectral analysis of short data records and density structure of the Earth.}",
    doi = "10.1103/PhysRevD.99.014013",
    journal = "Journal of Geophysics",
    volume = "42",
    number = "1",
    pages = "489",
    year = "1976"
}

@article{Zhang:2023bxs,
    author = "Zhang, Rui and Holligan, Jack and Ji, Xiangdong and Su, Yushan",
    title = "{Leading power accuracy in lattice calculations of parton distributions}",
    eprint = "2305.05212",
    archivePrefix = "arXiv",
    primaryClass = "hep-lat",
    doi = "10.1016/j.physletb.2023.138081",
    journal = "Phys. Lett. B",
    volume = "844",
    pages = "138081",
    year = "2023"
}

@article{Braun:2024snf,
    author = "Braun, Vladimir M. and Koller, Maria and Schoenleber, Jakob",
    title = "{Renormalons and power corrections in pseudo- and quasi-GPDs}",
    eprint = "2401.08012",
    archivePrefix = "arXiv",
    primaryClass = "hep-ph",
    doi = "10.1103/PhysRevD.109.074510",
    journal = "Phys. Rev. D",
    volume = "109",
    number = "7",
    pages = "074510",
    year = "2024"
}

@article{HadStruc:2021qdf,
    author = "Egerer, Colin and others",
    collaboration = "HadStruc",
    title = "{Transversity parton distribution function of the nucleon using the pseudodistribution approach}",
    eprint = "2111.01808",
    archivePrefix = "arXiv",
    primaryClass = "hep-lat",
    reportNumber = "JLAB-THY-21-3521",
    doi = "10.1103/PhysRevD.105.034507",
    journal = "Phys. Rev. D",
    volume = "105",
    number = "3",
    pages = "034507",
    year = "2022"
}

@article{Delmar:2023agv,
    author = "Delmar, Joseph and Alexandrou, Constantia and Cichy, Krzysztof and Constantinou, Martha and Hadjiyiannakou, Kyriakos",
    title = "{Gluon PDF of the proton using twisted mass fermions}",
    eprint = "2310.01389",
    archivePrefix = "arXiv",
    primaryClass = "hep-lat",
    doi = "10.1103/PhysRevD.108.094515",
    journal = "Phys. Rev. D",
    volume = "108",
    number = "9",
    pages = "094515",
    year = "2023"
}

@article{Fan:2022kcb,
    author = "Fan, Zhouyou and Good, William and Lin, Huey-Wen",
    title = "{Gluon parton distribution of the nucleon from (2+1+1)-flavor lattice QCD in the physical-continuum limit}",
    eprint = "2210.09985",
    archivePrefix = "arXiv",
    primaryClass = "hep-lat",
    reportNumber = "MSUHEP-22-033",
    doi = "10.1103/PhysRevD.108.014508",
    journal = "Phys. Rev. D",
    volume = "108",
    number = "1",
    pages = "014508",
    year = "2023"
}

@article{Braun:2018brg,
    author = "Braun, Vladimir M. and Vladimirov, Alexey and Zhang, Jian-Hui",
    title = "{Power corrections and renormalons in parton quasidistributions}",
    eprint = "1810.00048",
    archivePrefix = "arXiv",
    primaryClass = "hep-ph",
    doi = "10.1103/PhysRevD.99.014013",
    journal = "Phys. Rev. D",
    volume = "99",
    number = "1",
    pages = "014013",
    year = "2019"
}

@article{Liang:2019frk,
    author = "Liang, Jian and Draper, Terrence and Liu, Keh-Fei and Rothkopf, Alexander and Yang, Yi-Bo",
    collaboration = "XQCD",
    title = "{Towards the nucleon hadronic tensor from lattice QCD}",
    eprint = "1906.05312",
    archivePrefix = "arXiv",
    primaryClass = "hep-ph",
    doi = "10.1103/PhysRevD.101.114503",
    journal = "Phys. Rev. D",
    volume = "101",
    number = "11",
    pages = "114503",
    year = "2020"
}

@article{LatticePartonLPC:2021gpi,
    author = "Huo, Yi-Kai and others",
    collaboration = "Lattice Parton (LPC)",
    title = "{Self-renormalization of quasi-light-front correlators on the lattice}",
    eprint = "2103.02965",
    archivePrefix = "arXiv",
    primaryClass = "hep-lat",
    doi = "10.1016/j.nuclphysb.2021.115443",
    journal = "Nucl. Phys. B",
    volume = "969",
    pages = "115443",
    year = "2021"
}

@article{Liang:2017mye,
    author = "Liang, Jian and Liu, Keh-Fei and Yang, Yi-Bo",
    editor = "Della Morte, M. and Fritzsch, P. and G\'amiz S\'anchez, E. and Pena Ruano, C.",
    title = "{Lattice calculation of hadronic tensor of the nucleon}",
    eprint = "1710.11145",
    archivePrefix = "arXiv",
    primaryClass = "hep-lat",
    doi = "10.1051/epjconf/201817514014",
    journal = "EPJ Web Conf.",
    volume = "175",
    pages = "14014",
    year = "2018"
}

@phdthesis{Karpie:2019doj,
    author = "Karpie, Joseph Matheson",
    title = "{On the calculation of parton distributions from Lattice QCD}",
    reportNumber = "JLAB-THY-19-3077, DOE/OR/23177-4818",
    doi = "10.2172/1574126",
    school = "William-Mary Coll.",
    year = "2019"
}

@article{Zhang:2020dkn,
    author = "Zhang, Rui and Lin, Huey-Wen and Yoon, Boram",
    title = "{Probing nucleon strange and charm distributions with lattice QCD}",
    eprint = "2005.01124",
    archivePrefix = "arXiv",
    primaryClass = "hep-lat",
    reportNumber = "MSUHEP-20-008",
    doi = "10.1103/PhysRevD.104.094511",
    journal = "Phys. Rev. D",
    volume = "104",
    number = "9",
    pages = "094511",
    year = "2021"
}

@article{Good:2024iur,
    author = "Good, William and Hasan, Kinza and Lin, Huey-Wen",
    title = "{Toward the first gluon parton distribution from the LaMET}",
    eprint = "2409.02750",
    archivePrefix = "arXiv",
    primaryClass = "hep-lat",
    reportNumber = "MSUHEP-24-011",
    doi = "10.1088/1361-6471/ada815",
    journal = "J. Phys. G",
    volume = "52",
    number = "3",
    pages = "035105",
    year = "2025"
}

@article{Alexandrou:2021oih,
    author = "Alexandrou, Constantia and Constantinou, Martha and Hadjiyiannakou, Kyriakos and Jansen, Karl and Manigrasso, Floriano",
    title = "{Flavor decomposition of the nucleon unpolarized, helicity, and transversity parton distribution functions from lattice QCD simulations}",
    eprint = "2106.16065",
    archivePrefix = "arXiv",
    primaryClass = "hep-lat",
    doi = "10.1103/PhysRevD.104.054503",
    journal = "Phys. Rev. D",
    volume = "104",
    number = "5",
    pages = "054503",
    year = "2021"
}

@article{Li:2020xml,
    author = "Li, Zheng-Yang and Ma, Yan-Qing and Qiu, Jian-Wei",
    title = "{Extraction of Next-to-Next-to-Leading-Order Parton Distribution Functions from Lattice QCD Calculations}",
    eprint = "2006.12370",
    archivePrefix = "arXiv",
    primaryClass = "hep-ph",
    reportNumber = "JLAB-THY-20-3214",
    doi = "10.1103/PhysRevLett.126.072001",
    journal = "Phys. Rev. Lett.",
    volume = "126",
    number = "7",
    pages = "072001",
    year = "2021"
}

@article{Chen:2020ody,
    author = "Chen, Long-Bin and Wang, Wei and Zhu, Ruilin",
    title = "{Next-to-Next-to-Leading Order Calculation of Quasiparton Distribution Functions}",
    eprint = "2006.14825",
    archivePrefix = "arXiv",
    primaryClass = "hep-ph",
    doi = "10.1103/PhysRevLett.126.072002",
    journal = "Phys. Rev. Lett.",
    volume = "126",
    number = "7",
    pages = "072002",
    year = "2021"
}

@article{Izubuchi:2018srq,
    author = "Izubuchi, Taku and Ji, Xiangdong and Jin, Luchang and Stewart, Iain W. and Zhao, Yong",
    title = "{Factorization Theorem Relating Euclidean and Light-Cone Parton Distributions}",
    eprint = "1801.03917",
    archivePrefix = "arXiv",
    primaryClass = "hep-ph",
    reportNumber = "MIT-CTP-4960, MIT-CTP 4960",
    doi = "10.1103/PhysRevD.98.056004",
    journal = "Phys. Rev. D",
    volume = "98",
    number = "5",
    pages = "056004",
    year = "2018"
}

@article{Bhat:2022zrw,
    author = "Bhat, Manjunath and Chomicki, Wojciech and Cichy, Krzysztof and Constantinou, Martha and Green, Jeremy R. and Scapellato, Aurora",
    title = "{Continuum limit of parton distribution functions from the pseudodistribution approach on the lattice}",
    eprint = "2205.07585",
    archivePrefix = "arXiv",
    primaryClass = "hep-lat",
    doi = "10.1103/PhysRevD.106.054504",
    journal = "Phys. Rev. D",
    volume = "106",
    number = "5",
    pages = "054504",
    year = "2022"
}

@article{Zhang:2025hyo,
    author = "Zhang, Rui and Grebe, Anthony V. and Hackett, Daniel C. and Wagman, Michael L. and Zhao, Yong",
    title = "{Kinematically-enhanced interpolating operators for boosted hadrons}",
    eprint = "2501.00729",
    archivePrefix = "arXiv",
    primaryClass = "hep-lat",
    reportNumber = "FERMILAB-PUB-24-0968-T",
    journal = "",
    month = "1",
    year = "2025"
}

@article{Radyushkin:2017cyf,
    author = "Radyushkin, A. V.",
    title = "{Quasi-parton distribution functions, momentum distributions, and pseudo-parton distribution functions}",
    eprint = "1705.01488",
    archivePrefix = "arXiv",
    primaryClass = "hep-ph",
    reportNumber = "JLAB-THY-17-2455",
    doi = "10.1103/PhysRevD.96.034025",
    journal = "Phys. Rev. D",
    volume = "96",
    number = "3",
    pages = "034025",
    year = "2017"
}

@article{Ma:2017pxb,
      author         = "Ma, Yan-Qing and Qiu, Jian-Wei",
      title          = "{Exploring Partonic Structure of Hadrons Using ab initio
                        Lattice QCD Calculations}",
      journal        = "Phys. Rev. Lett.",
      volume         = "120",
      year           = "2018",
      number         = "2",
      pages          = "022003",
      doi            = "10.1103/PhysRevLett.120.022003",
      eprint         = "1709.03018",
      archivePrefix  = "arXiv",
      primaryClass   = "hep-ph",
      reportNumber   = "JLAB-THY-17-2542",
      SLACcitation   = "%%CITATION = ARXIV:1709.03018;%%"
}

@article{Braun:2007wv,
      author         = "Braun, V. and M{\"u}ller, Dieter",
      title          = "{Exclusive processes in position space and the pion
                        distribution amplitude}",
      journal        = "Eur. Phys. J.",
      volume         = "C55",
      year           = "2008",
      pages          = "349-361",
      doi            = "10.1140/epjc/s10052-008-0608-4",
      eprint         = "0709.1348",
      archivePrefix  = "arXiv",
      primaryClass   = "hep-ph",
      reportNumber   = "IPPP-07-54, DCPT-07-108",
      SLACcitation   = "%%CITATION = ARXIV:0709.1348;%%"
}

@article{Alexandrou:2020tqq,
    author = "Alexandrou, Constantia and Iannelli, Giovanni and Jansen, Karl and Manigrasso, Floriano",
    collaboration = "Extended Twisted Mass",
    title = "{Parton distribution functions from lattice QCD using Bayes-Gauss-Fourier transforms}",
    eprint = "2007.13800",
    archivePrefix = "arXiv",
    primaryClass = "hep-lat",
    doi = "10.1103/PhysRevD.102.094508",
    journal = "Phys. Rev. D",
    volume = "102",
    number = "9",
    pages = "094508",
    year = "2020"
}

@article{Bulava:2011yz,
    author = "Bulava, John and Donnellan, Michael and Sommer, Rainer",
    title = "{On the computation of hadron-to-hadron transition matrix elements in lattice QCD}",
    eprint = "1108.3774",
    archivePrefix = "arXiv",
    primaryClass = "hep-lat",
    reportNumber = "DESY-11-127, SFB-CPP-11-46",
    doi = "10.1007/JHEP01(2012)140",
    journal = "JHEP",
    volume = "01",
    pages = "140",
    year = "2012"
}

@article{Bali:2016lva,
    author = {Bali, Gunnar S. and Lang, Bernhard and Musch, Bernhard U. and Sch\"afer, Andreas},
    title = "{Novel quark smearing for hadrons with high momenta in lattice QCD}",
    eprint = "1602.05525",
    archivePrefix = "arXiv",
    primaryClass = "hep-lat",
    doi = "10.1103/PhysRevD.93.094515",
    journal = "Phys. Rev. D",
    volume = "93",
    number = "9",
    pages = "094515",
    year = "2016"
}

@article{Karpie:2019eiq,
    author = "Karpie, Joseph and Orginos, Kostas and Rothkopf, Alexander and Zafeiropoulos, Savvas",
    title = "{Reconstructing parton distribution functions from Ioffe time data: from Bayesian methods to Neural Networks}",
    eprint = "1901.05408",
    archivePrefix = "arXiv",
    primaryClass = "hep-lat",
    reportNumber = "JLAB-THY-19-2898",
    doi = "10.1007/JHEP04(2019)057",
    journal = "JHEP",
    volume = "04",
    pages = "057",
    year = "2019"
}

@article{Orginos:2017kos,
    author = "Orginos, Kostas and Radyushkin, Anatoly and Karpie, Joseph and Zafeiropoulos, Savvas",
    title = "{Lattice QCD exploration of parton pseudo-distribution functions}",
    eprint = "1706.05373",
    archivePrefix = "arXiv",
    primaryClass = "hep-ph",
    reportNumber = "JLAB-THY-17-2494",
    doi = "10.1103/PhysRevD.96.094503",
    journal = "Phys. Rev. D",
    volume = "96",
    number = "9",
    pages = "094503",
    year = "2017"
}

@article{Ji:2022ezo,
    author = "Ji, Xiangdong",
    title = "{Large-Momentum Effective Theory vs. Short-Distance Operator Expansion: Contrast and Complementarity}",
    eprint = "2209.09332",
    journal = "",
    archivePrefix = "arXiv",
    primaryClass = "hep-lat",
    month = "9",
    year = "2022"
}

@article{Ji:2024oka,
    author = "Ji, Xiangdong",
    title = "{Euclidean effective theory for partons in the spirit of Steven Weinberg}",
    eprint = "2408.03378",
    archivePrefix = "arXiv",
    primaryClass = "hep-ph",
    doi = "10.1016/j.nuclphysb.2024.116670",
    journal = "Nucl. Phys. B",
    volume = "1007",
    pages = "116670",
    year = "2024"
}

@article{Musch:2010ka,
    author = "Musch, Bernhard U. and Hagler, Philipp and Negele, John W. and Schafer, Andreas",
    title = "{Exploring quark transverse momentum distributions with lattice QCD}",
    eprint = "1011.1213",
    archivePrefix = "arXiv",
    primaryClass = "hep-lat",
    reportNumber = "MIT-CTP-4178, JLAB-THY-10-1266",
    doi = "10.1103/PhysRevD.83.094507",
    journal = "Phys. Rev. D",
    volume = "83",
    pages = "094507",
    year = "2011"
}

@article{Gao:2021dbh,
    author = "Gao, Xiang and Hanlon, Andrew D. and Mukherjee, Swagato and Petreczky, Peter and Scior, Philipp and Syritsyn, Sergey and Zhao, Yong",
    title = "{Lattice QCD Determination of the Bjorken-x Dependence of Parton Distribution Functions at Next-to-Next-to-Leading Order}",
    eprint = "2112.02208",
    archivePrefix = "arXiv",
    primaryClass = "hep-lat",
    doi = "10.1103/PhysRevLett.128.142003",
    journal = "Phys. Rev. Lett.",
    volume = "128",
    number = "14",
    pages = "142003",
    year = "2022"
}

@article{Ji:2020brr,
    author = {Ji, Xiangdong and Liu, Yizhuang and Sch\"afer, Andreas and Wang, Wei and Yang, Yi-Bo and Zhang, Jian-Hui and Zhao, Yong},
    title = "{A Hybrid Renormalization Scheme for Quasi Light-Front Correlations in Large-Momentum Effective Theory}",
    eprint = "2008.03886",
    archivePrefix = "arXiv",
    primaryClass = "hep-ph",
    doi = "10.1016/j.nuclphysb.2021.115311",
    journal = "Nucl. Phys. B",
    volume = "964",
    pages = "115311",
    year = "2021"
}

@article{Ji:2014gla,
    author = "Ji, Xiangdong",
    title = "{Parton Physics from Large-Momentum Effective Field Theory}",
    eprint = "1404.6680",
    archivePrefix = "arXiv",
    primaryClass = "hep-ph",
    doi = "10.1007/s11433-014-5492-3",
    journal = "Sci. China Phys. Mech. Astron.",
    volume = "57",
    pages = "1407--1412",
    year = "2014"
}

@article{Egerer:2021ymv,
    author = "Egerer, Colin and Edwards, Robert G. and Kallidonis, Christos and Orginos, Kostas and Radyushkin, Anatoly V. and Richards, David G. and Romero, Eloy and Zafeiropoulos, Savvas",
    collaboration = "HadStruc",
    title = "{Towards high-precision parton distributions from lattice QCD via distillation}",
    eprint = "2107.05199",
    archivePrefix = "arXiv",
    primaryClass = "hep-lat",
    reportNumber = "JLAB-THY-21-3457",
    doi = "10.1007/JHEP11(2021)148",
    journal = "JHEP",
    volume = "11",
    pages = "148",
    year = "2021"
}

@article{Candido:2024hjt,
    author = "Candido, Alessandro and Del Debbio, Luigi and Giani, Tommaso and Petrillo, Giacomo",
    title = "{Bayesian inference with Gaussian processes for the determination of parton distribution functions}",
    eprint = "2404.07573",
    archivePrefix = "arXiv",
    primaryClass = "hep-ph",
    doi = "10.1140/epjc/s10052-024-13100-1",
    journal = "Eur. Phys. J. C",
    volume = "84",
    number = "7",
    pages = "716",
    year = "2024"
}

@article{Hansen:2019idp,
    author = "Hansen, Martin and Lupo, Alessandro and Tantalo, Nazario",
    title = "{Extraction of spectral densities from lattice correlators}",
    eprint = "1903.06476",
    archivePrefix = "arXiv",
    primaryClass = "hep-lat",
    doi = "10.1103/PhysRevD.99.094508",
    journal = "Phys. Rev. D",
    volume = "99",
    number = "9",
    pages = "094508",
    year = "2019"
}

@inproceedings{Chen:2023zyy,
    author = "Chen, Jie and Edwards, Robert G. and Mao, Weizhen",
    title = "{Graph Contractions for Calculating Correlation Functions in Lattice QCD}",
    booktitle = "{Platform for Advanced Scientific Computing}",
    doi = "10.1145/3592979.3593409",
    year = "2023"
}

@article{Gao:2022uhg,
    author = "Gao, Xiang and Hanlon, Andrew D. and Holligan, Jack and Karthik, Nikhil and Mukherjee, Swagato and Petreczky, Peter and Syritsyn, Sergey and Zhao, Yong",
    title = "{Unpolarized proton PDF at NNLO from lattice QCD with physical quark masses}",
    eprint = "2212.12569",
    archivePrefix = "arXiv",
    primaryClass = "hep-lat",
    doi = "10.1103/PhysRevD.107.074509",
    journal = "Phys. Rev. D",
    volume = "107",
    number = "7",
    pages = "074509",
    year = "2023"
}

@article{BG,
    author = {Backus, George and Gilbert, Freeman},
    title = {The Resolving Power of Gross Earth Data},
    journal = {Geophysical Journal International},
    volume = {16},
    number = {2},
    pages = {169-205},
    year = {1968},
    month = {10},
    issn = {0956-540X},
    doi = {10.1111/j.1365-246X.1968.tb00216.x},
    url = {https://doi.org/10.1111/j.1365-246X.1968.tb00216.x},
    eprint = {https://academic.oup.com/gji/article-pdf/16/2/169/5891044/16-2-169.pdf},
}

@article{DelDebbio:2024lwm,
    author = "Del Debbio, Luigi and Lupo, Alessandro and Panero, Marco and Tantalo, Nazario",
    title = "{Bayesian solution to the inverse problem and its relation to Backus\textendash{}Gilbert methods}",
    eprint = "2409.04413",
    archivePrefix = "arXiv",
    primaryClass = "hep-lat",
    doi = "10.1140/epjc/s10052-025-13885-9",
    journal = "Eur. Phys. J. C",
    volume = "85",
    number = "2",
    pages = "185",
    year = "2025"
}

@article{Alexandrou:2021bbo,
    author = "Alexandrou, Constantia and Cichy, Krzysztof and Constantinou, Martha and Hadjiyiannakou, Kyriakos and Jansen, Karl and Scapellato, Aurora and Steffens, Fernanda",
    title = "{Transversity GPDs of the proton from lattice QCD}",
    eprint = "2108.10789",
    archivePrefix = "arXiv",
    primaryClass = "hep-lat",
    doi = "10.1103/PhysRevD.105.034501",
    journal = "Phys. Rev. D",
    volume = "105",
    number = "3",
    pages = "034501",
    year = "2022"
}

@article{Bhattacharya:2023nmv,
    author = "Bhattacharya, Shohini and Cichy, Krzysztof and Constantinou, Martha and Dodson, Jack and Metz, Andreas and Scapellato, Aurora and Steffens, Fernanda",
    title = "{Chiral-even axial twist-3 GPDs of the proton from lattice QCD}",
    eprint = "2306.05533",
    archivePrefix = "arXiv",
    primaryClass = "hep-lat",
    doi = "10.1103/PhysRevD.108.054501",
    journal = "Phys. Rev. D",
    volume = "108",
    number = "5",
    pages = "054501",
    year = "2023"
}

@article{Bhat:2020ktg,
    author = "Bhat, Manjunath and Cichy, Krzysztof and Constantinou, Martha and Scapellato, Aurora",
    title = "{Flavor nonsinglet parton distribution functions from lattice QCD at physical quark masses via the pseudodistribution approach}",
    eprint = "2005.02102",
    archivePrefix = "arXiv",
    primaryClass = "hep-lat",
    doi = "10.1103/PhysRevD.103.034510",
    journal = "Phys. Rev. D",
    volume = "103",
    number = "3",
    pages = "034510",
    year = "2021"
}

@incollection{QPhiX2,
author="Jo{\'o}, B{\'a}lint
and Kalamkar, Dhiraj D.
and Kurth, Thorsten
and Vaidyanathan, Karthikeyan
and Walden, Aaron",
editor="Taufer, Michela
and Mohr, Bernd
and Kunkel, Julian M.",
title={{Optimizing Wilson-Dirac Operator and Linear Solvers for Intel® KNL}},
booktitle="High Performance Computing: ISC High Performance 2016 International Workshops, ExaComm, E-MuCoCoS, HPC-IODC, IXPUG, IWOPH, $P^3$MA, VHPC, WOPSSS, Frankfurt, Germany, June 19--23, 2016, Revised Selected Papers",
year="2016",
publisher="Springer International Publishing",
address="Cham",
pages="415--427",
isbn="978-3-319-46079-6",
doi="10.1007/978-3-319-46079-6_30",
url="http://dx.doi.org/10.1007/978-3-319-46079-6_30"
}

@article{Zhang:2020rsx,
    author = {Zhang, Kuan and Li, Yuan-Yuan and Huo, Yi-Kai and Sch\"afer, Andreas and Sun, Peng and Yang, Yi-Bo},
    collaboration = "\ensuremath{\chi}QCD",
    title = "{RI/MOM renormalization of the parton quasidistribution functions in lattice regularization}",
    eprint = "2012.05448",
    archivePrefix = "arXiv",
    primaryClass = "hep-lat",
    doi = "10.1103/PhysRevD.104.074501",
    journal = "Phys. Rev. D",
    volume = "104",
    number = "7",
    pages = "074501",
    year = "2021"
}

@article{Dutrieux:2025axb,
    author = "Dutrieux, Herv{\'e} and Karpie, Joe and Monahan, Christopher J. and Orginos, Kostas and Radyushkin, Anatoly and Richards, David and Zafeiropoulos, Savvas",
    title = "{Comment on ''LaMET's Asymptotic Extrapolation vs. Inverse Problem''}",
    eprint = "2506.24037",
    archivePrefix = "arXiv",
    primaryClass = "hep-lat",
    reportNumber = "JLAB-THY-25-4392",
    month = "6",
    year = "2025",
    journal=""
}

@article{Alexandrou:2017huk,
    author = "Alexandrou, Constantia and Cichy, Krzysztof and Constantinou, Martha and Hadjiyiannakou, Kyriakos and Jansen, Karl and Panagopoulos, Haralambos and Steffens, Fernanda",
    title = "{A complete non-perturbative renormalization prescription for quasi-PDFs}",
    eprint = "1706.00265",
    archivePrefix = "arXiv",
    primaryClass = "hep-lat",
    reportNumber = "DESY-17-092",
    doi = "10.1016/j.nuclphysb.2017.08.012",
    journal = "Nucl. Phys. B",
    volume = "923",
    pages = "394--415",
    year = "2017"
}

@inproceedings{frontera,
author = {Stanzione, Dan and West, John and Evans, R. Todd and Minyard, Tommy and Ghattas, Omar and Panda, Dhabaleswar K.},
title = {Frontera: The Evolution of Leadership Computing at the National Science Foundation},
year = {2020},
isbn = {9781450366892},
publisher = {Association for Computing Machinery},
address = {New York, NY, USA},
url = {https://doi.org/10.1145/3311790.3396656},
doi = {10.1145/3311790.3396656},
booktitle = {Practice and Experience in Advanced Research Computing},
pages = {106–111},
numpages = {6},
location = {Portland, OR, USA},
series = {PEARC '20}
}

\end{document}